\documentclass[showpacs,floats,prb,aps,groupedaddress,showpacs,twocolumn,amsfonts,amssymb]{revtex4}
\usepackage{bm}
\usepackage{graphicx}

\newlength{\sepmod}
\begin{document}
\title{Monte Carlo study of the scaling of universal correlation lengths in
  three-dimensional O($n$) spin models}

\author{Martin Weigel}
\email{weigel@itp.uni-leipzig.de}

\author{Wolfhard Janke}
\email{janke@itp.uni-leipzig.de}

\affiliation{Institut f\"ur Theoretische Physik,
  Universit\"at Leipzig, Augustusplatz 10/11, 04109 Leipzig, Germany}

\date{July 21, 2000}

\begin{abstract}
  Using an elaborate set of simulational tools and statistically optimized methods of
  data analysis we investigate the scaling behavior of the correlation lengths of
  three-dimensional classical O($n$) spin models. Considering three-dimensional slabs
  $S^1\times S^1\times\mathbb{R}$, the results over a wide range of $n$ indicate the
  validity of special scaling relations involving universal amplitude ratios that are
  analogous to results of conformal field theory for two-dimensional systems. A
  striking mismatch of the $n\rightarrow\infty$ extrapolation of these simulations
  against analytical calculations is traced back to a breakdown of the identification
  of this limit with the spherical model.
\end{abstract}

\pacs{64.60.Fr, 75.10.Hk, 75.40.Mg, 11.25.Hf}

\maketitle

\section{Introduction}

The concept of scaling, the observation that singular observables vary in a
scale-free manner according to power laws when the driving parameter of a transition
(temperature, magnetic field, ...) is tuned towards a critical point, has since the
first observations been a key ingredient of the theory of critical phenomena
\cite{zinn-justin,kadanoff:domb}. Exploiting the symmetry of scale-invariance,
forming the geometrical basis for the power-law behavior in the vicinity of a
critical point, through the idea of real-space renormalization, scaling theory can be
mapped on the behavior of finite systems near the transition point of the bulk system
in the limit of diverging system sizes, the thermodynamic limit. This {\em
  finite-size scaling\/} (FSS) \cite{barber:72,fisher:74a,barber:domb} occurs with
scaling exponents generically linked to the exponents that govern scaling in the bulk
system. Thus the apparent weakness of finite system size that hampers simulational
approaches actually turns out to be their intrinsic strength, when exploring FSS
means exploring thermal scaling \cite{binder:schlad,bd:como}.

The significance of scaling theory for the understanding of critical phenomena
becomes quite exposed in the context of conformal field theory (CFT) for
two-dimensional systems \cite{belavin:84a}. In the course of exploiting the
additional invariances of conformal symmetry one is able to split the critical point
partition function of a lattice system into a sum over contributions from all the
scaling variables present in a specific model.  Consider a critical system on a
$L\times L'$ lattice with toroidal boundary conditions; then the partition function
decomposes as \cite{cardy:86a,cardy:86b}:
\begin{equation}
  Z(L,L') = e^{-fA+\pi c \delta/6}\sum_n{e^{-2\pi x_n\delta}},
\end{equation}
where $c$ is the central charge of the considered theory, $f$ the bulk free energy
per unit volume, $\delta=L'/L$, $A=LL'$, and the sum runs over the whole content of
scaling operators with dimensions $x_n$. Thus, the knowledge of the operator content
of a theory in connection with the corresponding scaling dimensions is equivalent to
an ``exact'' solution of the model on finite lattices.

\paragraph*{Two Dimensions --}
A particular example of a scaling relation in two dimensions that can be derived
assuming conformal invariance of critical point entities concerns the two-point
function in the limit of $L'\rightarrow\infty$.  It it generally sufficient to assume
translational, rotational, dilatational, and inversional invariance to imply
conformal invariance \cite{henkel:book}; homogeneity, isotropy and scale invariance
alone suffice to uniquely fix the critical, connected two-point function of an
operator $\phi$ in the infinite plane up to an overall normalization factor:
\begin{equation}
  \label{twopoint} {\langle \phi(z_1,\bar{z}_1) \phi(z_2,\bar{z}_2) \rangle}_c
  = (z_1-z_2)^{-x}(\bar{z}_1-\bar{z}_2)^{-x},
\end{equation} 
where $z_1$, $z_2$ are complex co-ordinates parametrizing the plane.
Then, one uses the logarithmic map
\begin{equation}
  w=\frac{L}{2 \pi} \ln z, \hspace{0.5cm} z \in \mathbb{C} \label{ln}
\label{logmap}
\end{equation}
to wrap the complex plane around an infinite length cylinder
$S^1\times\mathbb{R}$ of circumference $L$ with co-ordinates $w=u+iv$, where $v$
measures the polar angle along $S^1$ and $u$ the longitudinal direction along
$\mathbb{R}$. Assuming conformally covariant transformation behavior of the
(primary) operator $\phi$, one arrives at an expression for the two-point
function on the cylinder \cite{cardy:84a}:
\begin{eqnarray}
  {\langle \phi(w_1,\bar{w}_1)
  \phi(w_2,\bar{w}_2) \rangle}_c = {\left( \frac{2 \pi}{L} \right)}^{2x}
  {\left( \frac{\left| z_1 z_2 \right|}{{\left| z_1 - z_2 \right|}^2} \right)}^x
  \nonumber = \\
  {\left( \frac{2 \pi}{L} \right)}^{2x} {\left( 2\cosh {\frac{2 \pi}{L}
        (u_1-u_2)} -2\cos {\frac{2\pi}{L} (v_1-v_2)} \right)}^{-x}\!\!\!\!\!\!\!.\,\,\,\,
  \label{2Dkorr}
\end{eqnarray}
In the limit of large longitudinal distances $|u_1-u_2|\gg L$ and $v_1=v_2$, one is
left with a purely exponential drop with a correlation length
\begin{equation}
  \xi_{\parallel} = \frac{L}{2\pi x}. \label{xil}
\label{xi_par}
\end{equation}
Thus, utilization of conformal invariance yields a finite-size scaling relation {\em
  including the amplitude}, which is in contrast to renormalization group theory that
usually gives the scaling exponents and only certain amplitude ratios, but not the
amplitudes themselves. Since this result emerges from a field-theoretic description
of statistical mechanics that does not take into account the microscopical details of
the system, it is expected to be {\em universal\/}\cite{griffiths:70a}.  Note,
however, that this proposed universality goes beyond the usual notion of an universal
quantity and comprises three different aspects: (i) the correlation length of a given
operator should be the same within the associated universality class of models; (ii)
when looking at different operators, on the other hand, the form of Eq.\ 
(\ref{xi_par}) should be left unchanged, all operator-dependent information being
condensed in the scaling dimension $x$; (iii) finally, even when looking at models of
{\em different\/} universality classes, all that should change are the scaling
dimensions (and the definition of $\phi$), the validity of Eq.\ (\ref{xi_par}) being
untouched. Property (i) implies the ``hyperuniversality'' relation of Privman and
Fisher \cite{privman:84a}. In the following, we will refer to the whole extent of
aspects (i)-(iii) exceeding the usual notion of universality with the term
``hyperuniversal''. A corollary that is of importance for transfer matrix
calculations that use an unnormalized (quantum) Hamiltonian results from taking the
ratio of the correlation lengths of two primary operators, for example the densities
of magnetization and energy which are usually primary for spin models:
\begin{equation}
  \frac{\xi_\sigma}{\xi_\epsilon}=\frac{x_\epsilon}{x_\sigma}.
\label{ratio}
\end{equation}
Because of the independence from the overall amplitude $1/2\pi$ of Eq.\ 
(\ref{xi_par}) this relation might still stay valid when changing the geometry in a
way such that only this overall amplitude is altered. In terms of universality this
constitutes a weaker form of the aspect (i) above, namely universality of amplitude
ratios instead of amplitudes themselves; we will refer to this weaker property as
(i') in the following.

A suitable test-bed for these general field-theory results is, of course, given
by the exactly solvable two-dimensional Ising model. Using Eq.\ (\ref{xi_par})
and the generic relations between scaling dimensions and the conventional
critical exponents:
\begin{equation}
  \label{xdurchexp}
  x_\sigma =\frac{\beta}{\nu}, \hspace{1cm}
  x_\epsilon =\frac{1-\alpha}{\nu},
\end{equation}
giving $x_\sigma=1/8$ and $x_\epsilon=1$ for the two-dimensional Ising model, one
arrives at a ratio $x_\epsilon/x_\sigma=8$. A direct evaluation of the spin-spin
correlation length in the Onsager-Kaufman framework gives, as the leading term in the
scaling series, $\xi_\sigma=4L/\pi \equiv L/(2\pi\frac{1}{8})$, in agreement with the
CFT result \cite{domb:60,nightingale:76a,derrida:82a}. The same holds true for the
leading scaling amplitude of the energy-energy correlation function
\cite{bloete:83a}, $\xi_\epsilon=L/2\pi$.  Both amplitudes have also been evaluated
numerically to high precision in a Monte Carlo (MC) study \cite{diplom}, resulting in
perfect agreement with Eq.\ (\ref{xi_par}).

A possible alteration of the $S^1\times\mathbb{R}$ situation, namely changing the
boundary conditions along the $S^1$-direction from periodic to {\em antiperiodic\/}
has also been treated within the CFT framework, exploiting the fact that in the case
of the ferromagnetic nearest-neighbor Ising model the antiperiodic boundary
corresponds to the insertion of a seam of {\em anti\/}ferromagnetic bonds along this
boundary line.  This calculation yields \cite{cardy:84b,cardy:domb}:
\begin{equation}
  \xi_{\sigma} = \frac{4}{3 \pi} L, \hspace{1cm}
  \xi_{\epsilon} = \frac{1}{4 \pi} L,
\label{2Danti}
\end{equation}
again in good agreement with Monte Carlo data \cite{diplom}. Note, however, that this
last relation, in contrast to Eq.\ (\ref{xi_par}), is specific to the Ising model
{\em and\/} the special choice of the densities of magnetization and energy as
operators and thus is not ``hyperuniversal'' in the sense of properties (ii) and
(iii) presented above.

The amplitude-exponent relation Eq.\ (\ref{xi_par}) for two-dimensional systems has
been checked analytically or numerically and found valid for an impressive series of
further models like the Potts model and its percolation limit \cite{derrida:82a}, the
XY model \cite{luck:82a}, the symmetric eight-vertex model \cite{bloete:83a}, and
quantum spin models \cite{penson:84a} to name only the most prominent.

\paragraph*{Three Dimensions --}
On leaving the domain of two-dimensional systems towards higher dimensions, the
wealth of exact field theoretic calculations is instantly reduced to severe scarcity.
The conformal group coincides with the set of holomorphic functions in the special
case of spatial dimension $d=2$ and is thus infinite-dimensional as a group.  For
$d\ge 3$, unfortunately, it reduces to a simple Lie group with dimension $D\le
(d+1)(d+2)/2$ for any Riemannian, connected manifold. As a consequence, only in two
dimensions the postulate of conformal invariance is restrictive enough for a
classification of the operator contents of the different universality classes and
thus an exact solution of the critical theories within the limits of field-theory
assumptions. For $d\ge 3$, on the other hand, the implications of the
finite-dimensional conformal-group symmetry reach hardly beyond the consequences of
plain renormalization group theory exploiting dilatational invariance.  However,
since inversional symmetry is still present, a transformation like Eq.\ 
(\ref{logmap}) stays conformal in higher dimensions, now connecting the spaces
$\mathbb{R}^d$ and $S^{d-1}\times\mathbb{R}$. Applied to the two-point function one
arrives at a scaling relation analogous to Eq.\ (\ref{xi_par}), namely
$\xi_\parallel=R/x$, cp.\ Ref.\ \cite{cardy:85a}, which contains the $d=2$ result as
a special case assuming $L=2\pi R$, $R$ being the radius of $S^{d-1}$. Since
primarity of operators is {\em a priori\/} not well defined for $d\ge 3$, it is,
however, unclear for which operators this relation should hold. A numerical analysis
for this geometry, which has to cope with the fact that $S^{d-1}$ for $d\ge 3$ is a
truly curved space and thus hard to regularize by discrete lattices, will be
presented in a separate publication \cite{prep}.

On the other hand, the toroidal geometry $S^1\times\ldots\times S^1\times\mathbb{R}$,
which is much more convenient for numerical simulations, is not conformally flat and
thus no CFT predictions exist for this case. In spite of this theoretically
unfavorable situation a transfer matrix calculation for the Hamiltonian limit of the
three-dimensional Ising model on the geometry $S^1\times S^1\times\mathbb{R}\equiv
T^2\times\mathbb{R}$ by Henkel \cite{henkel:86a,henkel:87a,henkel:87b} rendered
results still comparable to the situation for the $S^{d-1}\times\mathbb{R}$ geometry.
For the ratios of leading scaling amplitudes of correlation lengths for
different boundary conditions (bc) he found
\begin{equation}
  \begin{array}{rcl}
    \xi_\sigma/\xi_\epsilon & = & 3.62(7) \hspace{0.5cm} \mbox{for periodic bc,} \\
    \xi_\sigma/\xi_\epsilon & = & 2.76(4) \hspace{0.5cm} \mbox{for antiperiodic bc.} \\
  \end{array}
\end{equation}
A comparison with the (inverse) ratio of the corresponding scaling dimensions,
\begin{equation}
  x_\epsilon/x_\sigma=\frac{(1-\alpha)/\nu}{\beta/\nu}=
  \frac{2(\nu d-1)}{\nu d-\gamma}=2.7264(13),
  \label{x_hyper}
\end{equation}
(cp.\ Table \ref{is_exp_tab} and Eq.\ (\ref{xdurchexp})) showed that even though the
original expectation to possibly find agreement in the case of periodic boundary
conditions as in the two-dimensional case was not met, the data are consistent with
the relation Eq.\ (\ref{ratio}) for the unorthodox case of {\em antiperiodic}
boundary conditions. Note that one has to compare {\em ratios\/} in this case,
because the quantum Hamiltonian used in the calculation is defined only up to on
overall normalization constant. This result is in qualitative agreement with a
Metropolis MC simulation by Weston \cite{weston:90}, who found ratios
$\xi_\sigma/\xi_\epsilon$ of about $3.7$ for periodic and $2.6$ for antiperiodic
boundary conditions, respectively.  Considering these striking observations it seems
interesting to check whether this behavior is just a coincidence or special feature
of the Ising model or instead indicates a general property of critical models on this
special three-dimensional geometry.

The rest of the paper is organized as follows. In Sec.\ II we introduce the general
class of models we want to examine and present the way we are going to discretize the
three-dimensional geometry $T^2\times\mathbb{R}$. We discuss simulation methods,
observables, estimators for measurements and parameters of the simulations. In Sec.\ 
III we outline the statistical tools used for the data analysis. It is quite hard to
extract high-precision information about correlation lengths from MC simulation data;
we will thus discuss the special path of data analysis we are going to proceed along
and present details of the statistical tools used there for.  This tool-set is
``calibrated'' with simulations of the {\em two-dimensional} Ising model, where exact
results for comparison are available. In Sec.\ IV we discuss the results for the
correlation lengths ratios of our simulations for the Ising, XY and (generalized)
Heisenberg models. Our results, already briefly announced in Ref.\ \cite{prl:99a},
confirm Henkel's findings on a high level of accuracy.  Furthermore this behavior
seems to carry through for the whole class of O($n$) spin models and is thus far from
being a ``numerical accident''. In Sec.\ V we try to rank our numerical findings in
the context of the classification of universality presented above. The type of the
model considered enters not only via a variation of the scaling dimensions, but also
influences the overall prefactor of Eq.\ (\ref{xi_par}). Sec.\ VI is devoted to the
discussion of the relation of our finite-$n$ results to the spherical model, which is
connected to the limit $n\rightarrow\infty$ of the class of O($n$) spin models. The
classic identification of both models seems to break down as soon as (multi-point)
correlation functions are considered. The final Sec.\ VII contains our conclusions.

\section{Models and Simulation}

Throughout this paper we consider classical, ferromagnetic, zero-field,
nearest-neighbor, O($n$) symmetric spin models with Hamiltonian
\begin{equation}
  \label{Hamilton}        
  {\cal H} = -J \sum_{\langle ij\rangle} \bm{\sigma}_i \cdot \bm{\sigma}_j,\;\;
  \bm{\sigma}_i \in S^{n-1}.
\end{equation}
The underlying lattice is taken to be simple cubic with dimensions $L_x\times
L_y\times L_z$. Special cases of this class of models include the Ising ($n=1$), XY
($n=2$), and Heisenberg ($n=3$) models.  This Hamiltonian has the advantage of
representing a whole class of models with critical points in three dimensions, tuned
by the single parameter $n$. According to the $T^2\times\mathbb{R}$ geometry we set
$L_x=L_y$ and apply periodic {\em or\/} antiperiodic boundary conditions in the $x$
and $y$ directions.  In both cases we use periodic boundary conditions in the
$z$-direction to eliminate surface effects that are also absent in the
$L_z\rightarrow\infty$ case assumed in Eq.\ (\ref{2Dkorr}). To reduce effects of
finite size in $z$-direction one has to ensure that $L_z\gg \xi_\parallel$, a
concrete rule will be given below.

\begin{table}[bt]
  \caption{Literature estimates for the critical exponents $\nu$ and $\gamma$ of the
    three-dimensional Ising model.\label{is_exp_tab}}
  \begin{center}
    \begin{tabular*}{\linewidth}{@{\extracolsep{\fill}}lll}
      \toprule
      Method & \multicolumn{1}{c}{$\nu$} & \multicolumn{1}{c}{$\gamma$} \\ \colrule  
      $g$-expansion\cite{guida:98a} & 0.6304(13) & 1.2396(13) \\
      $\epsilon$-expansion\cite{guida:98a} & 0.6305(25) & 1.2380(50) \\
      series\cite{butera:97a} & 0.6315(8) & 1.2388(10) \\
      series\cite{pelissetto:99a} & 0.63002(23) & 1.2371(4) \\
      MC\cite{ferrenberg:91a} & 0.6289(8) & 1.239(7) \\
      MC\cite{bloete:95a} & 0.6301(8) & 1.237(2) \\
      MC\cite{bloete:99a} & 0.6303(6) & 1.2372(13) \\
      MC\cite{hasenbusch:99b} & 0.6298(5) & 1.2365(5) \\ \colrule
      weighted mean & 0.63005(18) & 1.23717(28) \\ \botrule
    \end{tabular*}
  \end{center}
\end{table}

\begin{figure*}[t]
  \mbox{ \raisebox{5.5ex}{(a)} \hspace{1ex} \includegraphics[clip=true]{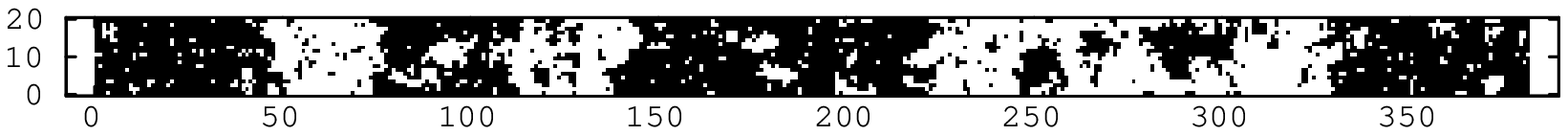}}
  \mbox{ \raisebox{5.5ex}{(b)} \hspace{1ex} \includegraphics[clip=true]{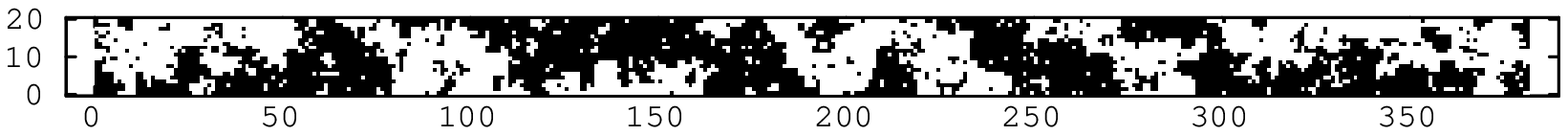}}
  \caption{Typical spin configurations for the two-dimensional Ising model on strips of size
    $20\times 382$. (a) periodic boundary conditions; (b) antiperiodic boundary
    conditions. Note that the visible geometric clusters differ from the stochastic
    clusters of the cluster update algorithm.}
  \label{snaps}
\end{figure*}

In view of the problem of critical slowing down, we use the Wolff single cluster
update algorithm\cite{wolff:89a} for all O($n$) model simulations, cp.\ 
\cite{prl:99a}. The adaption of this update procedure to the case of antiperiodic
boundary conditions along the torus directions is straightforward if one exploits the
above mentioned equivalence of an antiperiodic boundary to the insertion of a seam of
antiferromagnetic bonds along the boundary line for the case of nearest-neighbor
interactions. Considering the Ising model or, alternatively, embedded Ising spins for
$n>1$ models \cite{wolff:90a}, this means: adjacent spins interacting
antiferromagnetically are connected with a bond obeying the Swendsen-Wang probability
$p=1-\exp(-2\beta J)$ in case of {\em opposite\/} orientation and are left unbonded
in case of identical orientation. This rule exactly reflects the change in energy
compared to the ferromagnetic case and thus trivially satisfies detailed balance.

The main observables of our simulations are the connected correlation functions of
the densities of magnetization and energy:
\begin{equation}
  \begin{array}{rcl}
    G_{\sigma}^c({\bf x}_1,{\bf x}_2) & = & \langle\bm{\sigma}({\bf x}_1)\cdot
    \bm{\sigma}({\bf x}_2)\rangle-\langle\bm{\sigma}\rangle\cdot\langle\bm{\sigma}\rangle, \\
    G_{\epsilon}^c({\bf x}_1,{\bf x}_2) & = & \langle\epsilon({\bf x}_1)\,
    \epsilon({\bf x}_2)\rangle-\langle\epsilon\rangle\langle\epsilon\rangle. \\
  \end{array}
  \label{conncorr}
\end{equation}
We define the energy density as a local sum over the nearest neighborhood ${\bf x}'$
of a spin ${\bf x}$ (${\bf x}'\,\mathrm{nn}\,{\bf x}$):
\begin{equation}
  \epsilon({\bf x})=-\frac{J}{2}\sum_{{\bf x}'\,\mathrm{nn}\,{\bf x}}\bm{\sigma}({\bf x})
  \cdot\bm{\sigma}({\bf x'}),
  \label{e_dens}
\end{equation}
the factor $1/2$ ensuring that $E=\sum_{\bf x}\epsilon({\bf x})$.  It is
straightforward to construct a bias-reduced estimator for the case of $({\bf x}_2 -
{\bf x}_1)\parallel\hat{e}_z$, corresponding to the correlation length
$\xi=\xi_\parallel$: first, taking advantage of the translation invariance of the
systems along the $z$-axis established by a periodic boundary, one can average over
the ``layers'' $i\equiv|z_2-z_1| = \mbox{const}$. To improve on that
consider a ``zero-mode projection'' \cite{wj:93a}, i.e.\ define layered variables
\begin{equation}
  \bar{\cal{O}}_t(z) = \frac{1}{L_xL_y}\sum_{{\bf x}', z'=z}{\cal O}_t({\bf x}'),
\end{equation}
where ${\cal{O}}_t=\bm{\sigma}_t$ or $\epsilon_t$ denotes the times series of MC
measurements, and consider the estimator
\begin{eqnarray}
  \hat{G}^{c,\parallel}_{\cal O}(i) & = &
  \frac{1}{T}\sum_{t=1}^T\frac{1}{L_z}\sum_{|z_2-z_1|=i}\bar{{\cal O}}_t(z_1)\bar{\cal
    O}_t(z_2)  \nonumber \\
  & & -{\left(\frac{1}{TL_z}\sum_{t=1}^T\sum_z\bar{\cal O}_t(z)\right)}^2,
\label{Gest}
\end{eqnarray}
where $T$ denotes the length of the MC time series. This estimator obviously does not
directly measure $G^{c,\parallel}$, but inspecting the continuum form Eq.\ 
(\ref{2Dkorr}) reveals that the deviation stemming from transversal
cross-correlations entering the estimator declines exponentially with increasing
longitudinal distance $i$ and thus becomes irrelevant for the long-distance behavior
we are interested in. Numerical investigations confirm that these considerations stay
correct when passing to three dimensions \cite{diplom}. In the large-distance regime
zero-mode projection reduces the variance of correlation function estimates by a
factor inversely proportional to the layer volume $L_xL_y$. Note that the given
estimator for the disconnected part ${\langle{\cal O}\rangle}^2$ has a bias that
vanishes as $1/T$ in the large-$T$ limit.

As mentioned above, periodic boundary conditions in $z$-direction eliminate surface
effects associated with this direction, but still effects of finite $L_z$ will
trigger deviations from the $L_z\rightarrow\infty$ limit assumed in Eq.\ 
(\ref{xi_par}). Inspecting the form of Eq.\ (\ref{2Dkorr}) in the limit of distances
$i\gg \xi_\parallel$ one expects longitudinal correlations according to:
\begin{equation}
  G^{c,\parallel}(i) \propto e^{-i/\xi_\parallel} + e^{-(L_z-i)/\xi_\parallel},
\end{equation}
i.e.\ the exponential decay is superimposed by an exponentially increasing part.
Thus, using too small values of $L_z$ results in an effective underestimation of
correlation lengths. In order to satisfy $L_z\gg\xi_\parallel$ in a systematic way,
i.e.\ to keep this effect away from the region of clear signal for measuring the
correlation lengths, and assuming linear scaling of correlation lengths according to
$\xi_\parallel=AL_x$, one has to keep the ratio $L_z/\xi_\parallel=L_z/AL_x$ fixed
and therefore has to scale $L_z$ proportionally to $L_x$.  Simulations for the case
of the {\em two-dimensional} Ising model show that these finite-size effects are
negligible compared to the statistical errors for $L_z/\xi_\parallel\gtrsim 10$ and
lengths of time series of about $10^6$ to $10^7$ measurements \cite{diplom}. Adding a
safety margin the longitudinal system sizes for the simulations in three dimensions
where chosen such that $L_z/\xi_\parallel\approx 15$, the scaling amplitude $A$ being
estimated from a simulation of an ``oversized'' system. Since
$\xi_\sigma>\xi_\epsilon$ for all models under consideration, the amplitude $A_\sigma$
of the spin-spin correlation length scaling is significant for the satisfaction of
this condition. Note that from Eq.\ (\ref{Gest}) increasing $L_z$ also has the
positive side effect of improving the statistics of the correlation function
estimation.

In order to judge the efficiency of the used cluster update algorithm and to ensure
reasonable usage of computer time we evaluated integrated autocorrelation times
$\tau_{\rm int}$ using a binning technique \cite{flyvbjerg:89a}. The strong asymmetry
of the model lattices reduces the average size of clusters and thus Wolff's cluster
update algorithm does not perform as good as on (hyper-)cubic lattices, resulting in
increased autocorrelation times. Since measurements of $\hat{G}^{c,\parallel}$ are
computationally expensive compared to update steps, but the statistical gain vanishes
with increasing $\tau_\mathrm{int}$, measurements were done with frequencies of about
$1/\tau_\mathrm{int}$. Approaching the low-temperature phase, antiperiodic boundary
conditions in the torus directions produce a spatially stable boundary of the
geometric clusters along the antiferromagnetic seam, which in turn enforces a second
boundary along the $z$ direction. This results in a further reduction of the average
cluster size compared to the periodic boundary case. Fig.\ \ref{snaps} shows typical
configurations for the case of the (two-dimensional) Ising model.

\section{Data Analysis}

\begin{figure}
  \includegraphics[angle=-90,clip=true,keepaspectratio=true,width=\linewidth]{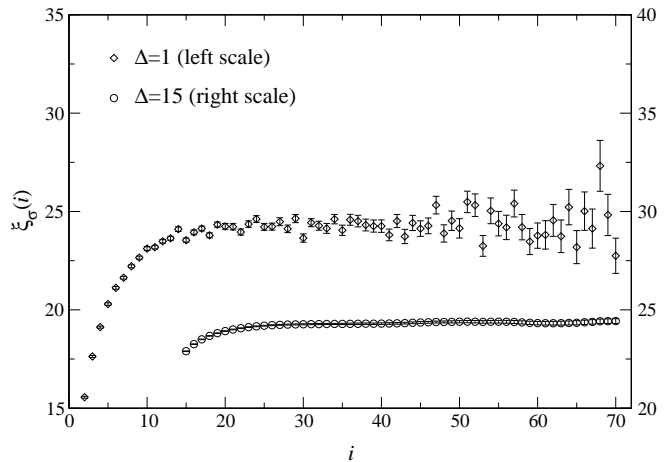} 
  \caption{
    Correlation length estimates according to Eq.\ (\ref{diffmethoddelta}) and ${\cal
      O}=\bm{\sigma}$ for a $30^2\times382$ Ising system with periodic boundary
    conditions for two choices of the typical distance $\Delta$. The plateau regimes
    collapse if both ordinates are scaled identically.}
  \label{xideltaplot}
\end{figure}

Having sampled correlation functions according to Eq.\ (\ref{Gest}) and assuming the
functional form $G^{c,\parallel}(i)=a\exp{(-i/\xi_\parallel)}+b$, we refrain from
using instrinsically unstable non-linear three-parameter fits and resort to the
following estimator instead,
\begin{equation}
  \hat{\xi}_{\cal O}(i)=\Delta{\left[\ln\frac{\hat{G}^{c,\parallel}_{\cal O}(i)
        -\hat{G}^{c,\parallel}_{\cal O}(i-\Delta)}
      {\hat{G}^{c,\parallel}_{\cal O}(i+\Delta)-
        \hat{G}^{c,\parallel}_{\cal O}(i)}\right]}^{-1},
  \label{diffmethoddelta}
\end{equation}
which eliminates the additive and multiplicative constants $a$ and $b$ above. Note
that it is not allowed to assume $b=0$ {\em a priori\/} for time series of finite
length, cp.\ Ref.\ \cite{prl:99a}. Apart from stability considerations this approach
allows for computational simplifications, since correlation functions can be sampled
irrespective of normalization and the biased estimation of the disconnected part
${\langle{\cal O}\rangle}^2$ can be dropped. In addition, Eq.\ 
(\ref{diffmethoddelta}) simplifies the distinction of the long-distance part of the
correlation function from the short-distance region: as the explicit two-dimensional
expression Eq.\ (\ref{2Dkorr}) implies, exponential decay will only occur
asymptotically, but with deviations decaying themselves exponentially; apart from
that, lattice artefacts that are not reflected in the continuum expression Eq.\ 
(\ref{2Dkorr}) additionally distort the short-distance behavior. Fig.\ 
\ref{xideltaplot} shows an example plot of the spin-spin correlation length estimates
$\hat{\xi}_\sigma(i)$ for the Ising model.  The transition from the short-distance
region that should not be used for the final estimate to the purely exponential
long-distance behavior is clearly visible.  The parameter $\Delta$ in Eq.\ 
(\ref{diffmethoddelta}) can be used to tune the signal-noise ratio for the
correlation length estimate; increasing $\Delta$ dramatically reduces the apparent
statistical fluctuations in $\hat{\xi}(i)$, cp.\ Fig.\ \ref{xideltaplot}. Note,
however, that the reduction of variances for individual distances $i$ is accompanied
by an increase of cross-correlations between estimates for adjacent estimates, so
that the error of an {\em average\/} over a region of distances becomes minimal for a
value $\Delta$ clearly below its allowed maximum. As a compromise, we use
$\Delta\approx 2\xi_\epsilon$ for both estimators $\hat{\xi}_\sigma(i)$ and
$\hat{\xi}_\epsilon(i)$.

Naive estimates for the statistical errors (variances) of complex, non-linear
combinations of observable measurements like the estimator Eq.\ 
(\ref{diffmethoddelta}) are extremely biased due to two effects: even for quite
sparse measurements with frequencies around $1/\tau_\mathrm{int}$ successive elements
of the time series are still correlated, generically leading to systematic
underestimation of variances. This effect is being eliminated by the grouping
together of measurements to sub-averages of length $\mu$ (``binning'')
\cite{flyvbjerg:89a}, which leads to an asymptotically uncorrelated time series of
length $T'=T/\mu$ used in the further process of error estimation. For the
production-run time-series the bin size was chosen to regularly include several
thousand measurements, which is far in the asymptotical regime. Secondly, the strong
non-linearity of estimators like Eq.\ (\ref{diffmethoddelta}) forbids the use of the
usual formula for the standard deviation of a set of measurements. A common solution
to this problem is the use of the Gaussian error propagation formula, which, however,
only uses a lowest order Taylor series approximation to the functions and assumes
Gaussian distribution of the mean values, i.e.\ long enough time series for all
observables. A far more general ansatz is given by resampling techniques such as the
``jackknife'' \cite{efron:82} that apply to a quite general set of probability
distributions and capture function non-linearities exactly. The jackknife variance
and bias estimators mimic the brute force error estimation method of comparing $k$
completely independent MC time series of lengths $T'$ and applying the naive
estimates: removing single elements (i.e.\ bins) of a single time series of length
$T'$ one by one results in $T'$ time series of length $T'-1$, e.g.\ for the
correlation function estimates:
\begin{equation}
  \hat{G}_{(s)}(i) = \frac{1}{T'-1}\sum_{t\neq s}\hat{G}_t(i),
\end{equation}
resulting in jackknife-block estimates for the correlation length of:
\begin{equation}
  \begin{array}{rcl}
    {\displaystyle\hat{\xi}_{(s)}(i)} & = & \Delta
    {\displaystyle{\left[\ln\frac{\hat{G}_{(s)}(i)-\hat{G}_{(s)}(i-\Delta)}
          {\hat{G}_{(s)}(i+\Delta)-\hat{G}_{s}(i)}\right]}^{-1},}\vspace{0.3cm}\\
    {\displaystyle\hat{\xi}_{(\cdot)}(i)} & = &
    {\displaystyle\frac{1}{T'}\sum_{s}\hat{\xi}_{(s)}(i).} \\
  \end{array}
\end{equation}
Then the jackknife estimate of variance is given by:
\begin{equation}
  \widehat{\mathrm{VAR}}(\hat{\xi}(i))=\frac{T'-1}{T'}\sum_{s=1}^{T'}
  {\left(\hat{\xi}_{(s)}(i)-\hat{\xi}_{(\cdot)}(i)\right)}^2.
\end{equation}
Note the missing factor of $1/(T'-1)^2$ as compared to the usual variance estimate
which accounts for the trivial correlation between the $T'$ jackknife-block estimates.
One can show that this estimator, apart from the reweighting prefactor $(T'-1)/T'$, is
strictly conservative, i.e.\ deviations from the true variance are always positive
\cite{efron:82}.  Similarly, the resampling scheme provides an estimate for the bias
of estimators, namely:
\begin{equation}
  \widehat{\mathrm{BIAS}}(\hat{\xi}(i))=(T'-1)(\hat{\xi}_{(\cdot)}(i)-\hat{\xi}(i)),
  \label{xitild}
\end{equation}
and thus offers a bias corrected correlation length estimate as
$\tilde{\xi}(i)=T'\hat{\xi}(i)-(T'-1)\hat{\xi}_{(\cdot)}(i)$. Since in
non-pathological cases the bias of an estimator vanishes with increasing length of
the time series, the jackknife bias estimate provides a good check for whether the
considered series are long enough to neglect bias. A jackknife error estimate for
these bias-corrected estimators is possible iterating the jackknife resampling scheme
to second order \cite{berg:92a}.

Since Eq.\ (\ref{diffmethoddelta}) gives a vector of estimators for the correlation
length instead of only a single one, an improved final estimate can be achieved by an
average over the $\hat{\xi}(i)$. However, as for example Fig.\ \ref{xideltaplot}
reveals, only a certain range of distances $i=i_\mathrm{min},\ldots,i_\mathrm{max}$
is suited for this purpose, where the lower bound $i_\mathrm{min}$ results mainly
from small-distance deviations as reflected by Eq.\ (\ref{2Dkorr}), whereas the large
distance bound $i_\mathrm{max}$ cuts off the region where the signal of exponential
fall-off drops below the size of statistical fluctuations, so that error estimates
become inaccurate and eventually the estimator Eq.\ (\ref{diffmethoddelta}) becomes
maldefined due to negative arguments of the logarithm.  Conventionally, averaging
over the estimates $\hat{\xi}(i)$ for $i=i_{\rm min},\ldots,i_{\rm max}$ would be
done with weights $\alpha_i \propto 1/\sigma^2(\hat{\xi}(i))$ that minimize the
theoretical variance of the mean value.  This prescription, however, neglects
correlations between the individual estimates.  Note that cross-correlations between
adjacent estimates $\hat{\xi}(i)$ are quite large, not only because large scale
fluctuations of the correlation functions are dominant, but also since the used
estimator Eq.\ (\ref{diffmethoddelta}) explicitly introduces such correlations
increasing in range with increasing $\Delta$. As a simple variational calculation
shows, for the case of correlated variables to be averaged over, one has to choose
the weights according as
\begin{equation}
  \alpha_k=\frac{\sum_i(\Gamma^{-1})_{ik}}{\sum_{i,j}(\Gamma^{-1})_{ij}},
  \label{alphas}
\end{equation}
in order to minimize the variance of the mean value. Here, $\Gamma \in
\mathbb{R}_{p\times p},\,\,p=i_\mathrm{max}-i_\mathrm{min}+1$, denotes the covariance
matrix of the $\hat{\xi}(i)$. $\Gamma$ itself can be estimated within the jackknife
resampling scheme as:
\begin{eqnarray}
  \widehat{\mathrm{CORR}}_{ij} \equiv \widehat{\mathrm{CORR}}(\hat{\xi}(i),\hat{\xi}(j)) = \nonumber \\
  = \frac{T'-1}{T'}\sum_{s=1}^{T'}{\left(\hat{\xi}_{(s)}(i)-\hat{\xi}_{(\cdot)}(i)\right)}
  {\left(\hat{\xi}_{(s)}(j)-\hat{\xi}_{(\cdot)}(j)\right)}.
\end{eqnarray} 
The fact that, considering Eq.\ (\ref{alphas}), variance and covariance estimates
directly influence the final results for the correlation lengths, gave the motivation
for the quite careful statistical treatment presented above.

\begin{figure}[tb]
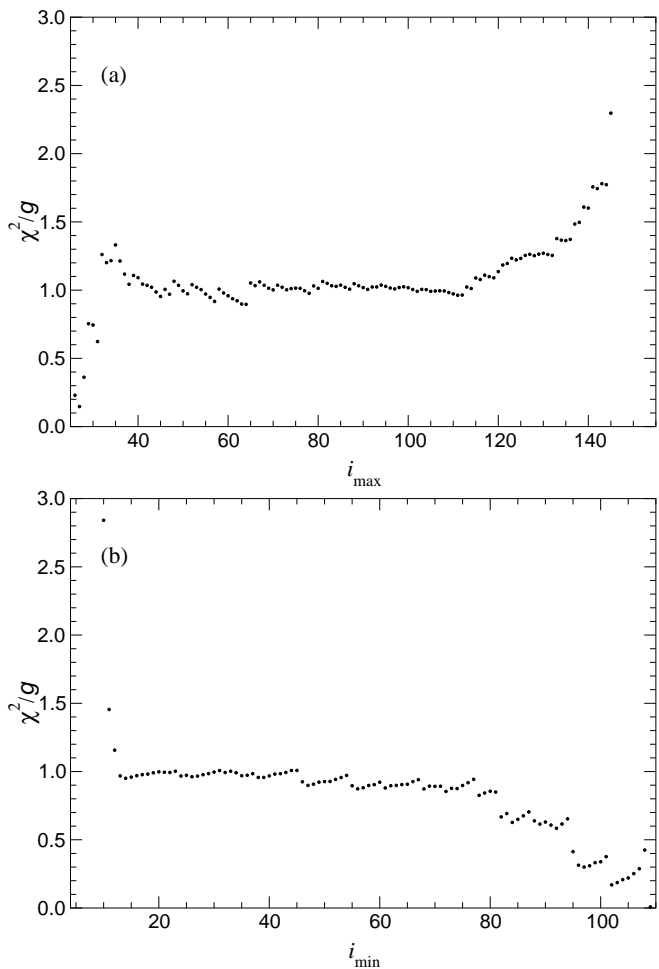

  \includegraphics[angle=-90,width=\linewidth,clip=true,keepaspectratio=true]{chi2sschnitt1.eps}
  \includegraphics[angle=-90,width=\linewidth,clip=true,keepaspectratio=true]{chi2sschnitt2.eps}
  \caption{Sections of $\hat{\chi}^2/g(i_\mathrm{min},i_\mathrm{max})$ for the spin
    correlation length of an Ising system. (a)
    $\{\hat{\chi}^2/g\,|\,i_\mathrm{min}=25\}$; (b)
    $\{\hat{\chi}^2/g\,|\,i_\mathrm{max}=110\}$. The ``wavy'' structure results from
    $\Delta=4$ in Eq.\ (\ref{diffmethoddelta}).}
  \label{chi2plot}
\end{figure}

Finally, the selection of the regime $i=i_\mathrm{min},\ldots,i_\mathrm{max}$ can,
besides the obvious eyeball method, also be done in a way based on statistical
criteria.  Interpreting the average over the $\hat{\xi}(i)$ as a {\em fit\/} of the
estimated $\hat{\xi}(i)$ values to the trivial function
$f(\hat{\xi})=\bar{\xi}=\mathrm{const}$, the systematic deviations from the plateau
regime for very small and very large distances $i$ should be clearly reflected in
quality-of-fit parameters. Thus, looking at the $\chi^2$-distribution,
\begin{equation}
  \hat{\chi}^2=\sum_{i,j=i_\mathrm{min}}^{i_\mathrm{max}}(\hat{\xi}(i)-\bar{\xi})
  (\hat{\Gamma}^{-1})_{ij}(\hat{\xi}(j)-\bar{\xi}),
  \label{chi2}
\end{equation}
will be a good criterion for judging the ``flatness'' of the plateau regime
$i_\mathrm{min},\ldots,i_\mathrm{max}$ included in the average. Again, as an
estimator $\hat{\Gamma}_{ij}$ for the covariance matrix one can use the jackknife
expression $\widehat{\mathrm{CORR}}_{ij}$.  Then finding the optimal region of
distances for the average is equivalent to the optimization problem
$|\hat{\chi}^2/g-1|\rightarrow\mathrm{min}$, with
$g=i_\mathrm{max}-i_\mathrm{min}=p-1$ denoting the number of degrees of freedom of
the fit. However, this ansatz of optimization bears some uncertainties: minimizing
the distance of $\hat{\chi}^2/g$ from $1$ supposes that the optimal choice includes
estimates $\hat{\xi}(i)$ whose dispersion around $\bar{\xi}$ is exactly reflected by
the estimated variances. In view of the jackknife's tendency to overestimate errors
it might be more favorable to minimize $|\hat{\chi}^2/g|$ itself. Furthermore,
considering the statistical nature of the data, the absolute minimum of
$|\hat{\chi}^2/g-1|$ or $|\hat{\chi}^2/g|$ sometimes happens to be an isolated
fluctuation, far apart from the bulk of next-to-optimal solutions. Finally, this
optimization procedure tends to result in minimal values for very small regime sizes
$p$ since the fit becomes trivial for very small numbers of points; this, however,
conflicts with another possible goal of optimization, namely the minimization of the
overall variance of the final result. To circumvent these problems we resort to
considering the whole two-dimensional distribution
$\hat{\chi}^2/g(i_\mathrm{min},i_\mathrm{max})$. It is characterized by a rather flat
plateau regime for intermediate values of $i_\mathrm{min}$ and $i_\mathrm{max}$ and
steep increases at the boundaries, cp.\ Fig.\ \ref{chi2plot}. A good recipe for the
determination of bounds is then given by first choosing a preliminary
$i_\mathrm{min}$ well above the steep ascent for small $i$; then a plot like Fig.\ 
\ref{chi2plot}(a) allows to determine the upper bound $i_\mathrm{max}$. Finally, a
plot of $\{\hat{\chi}^2/g\,|\,i_\mathrm{max}=\mathrm{const}\}$ determines the final
lower bound $i_\mathrm{min}$, cp.\ Fig.\ \ref{chi2plot}(b).

\begin{figure}[tb]
  \begin{center}
    \includegraphics[angle=-90,clip=true,width=\linewidth,keepaspectratio=true]{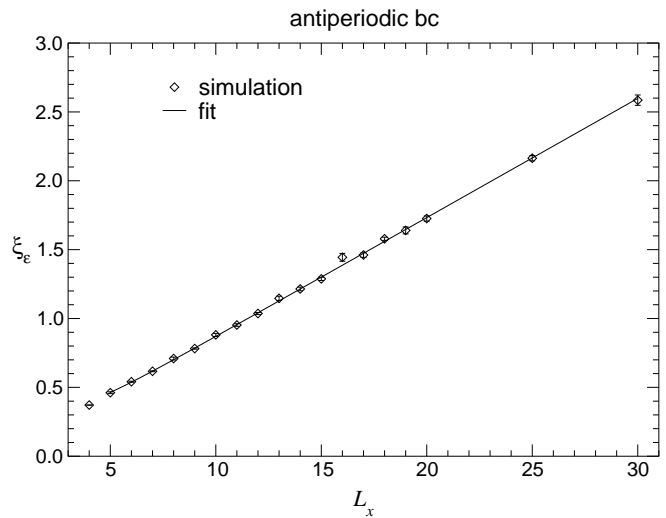}
  \end{center}
  \caption{Finite-size scaling of the energy-energy correlation length of the
    three-dimensional Ising model with antiperiodic boundary conditions. The other
    scaling plots look similar; we show the worst case. The fit was done to the
    functional form Eq.\ (\ref{fitform}).\label{scale_ex}}
\end{figure}

To test the methods of data analysis described in this section we performed
simulations of the {\em two-dimensional\/} Ising model. Using a series of systems
with $L_x=5,\ldots,20$ and finite-size scaling fits including an effective
higher-order correction term of the form $\xi(L_x)=AL_x+BL_x^{\kappa}$, we find for
the leading correlation lengths scaling amplitudes $A_{\sigma/\epsilon}$ final
estimates for the case of periodic boundary conditions of $A_\sigma=1.27374(81)$ and
$A_\epsilon=0.1583(17)$, in excellent agreement with the exact results
$A_\sigma=4/\pi\approx1.27324$ and $A_\epsilon=1/2\pi\approx0.15915$, cp.\ Eq.\ 
(\ref{xi_par}). For the case of antiperiodic boundary conditions we arrive at
$A_\sigma=0.42410(30)$ and $A_\epsilon=0.07984(38)$, compared to CFT results of
$A_\sigma=4/3\pi\approx0.42441$ and $A_\epsilon=1/4\pi\approx0.07958$, cp.\ Eq.\ 
(\ref{2Danti}).

\section{Results: amplitude ratios}

Let us now turn to the three-dimensional geometry $T^2\times\mathbb{R}$ and the
determination of amplitude ratios according to Eq.\ (\ref{ratio}). We report the
results of simulations for the O($n$) spin models for $n=1$, $2$, $3$, and $10$.

\begin{figure*}[tb]
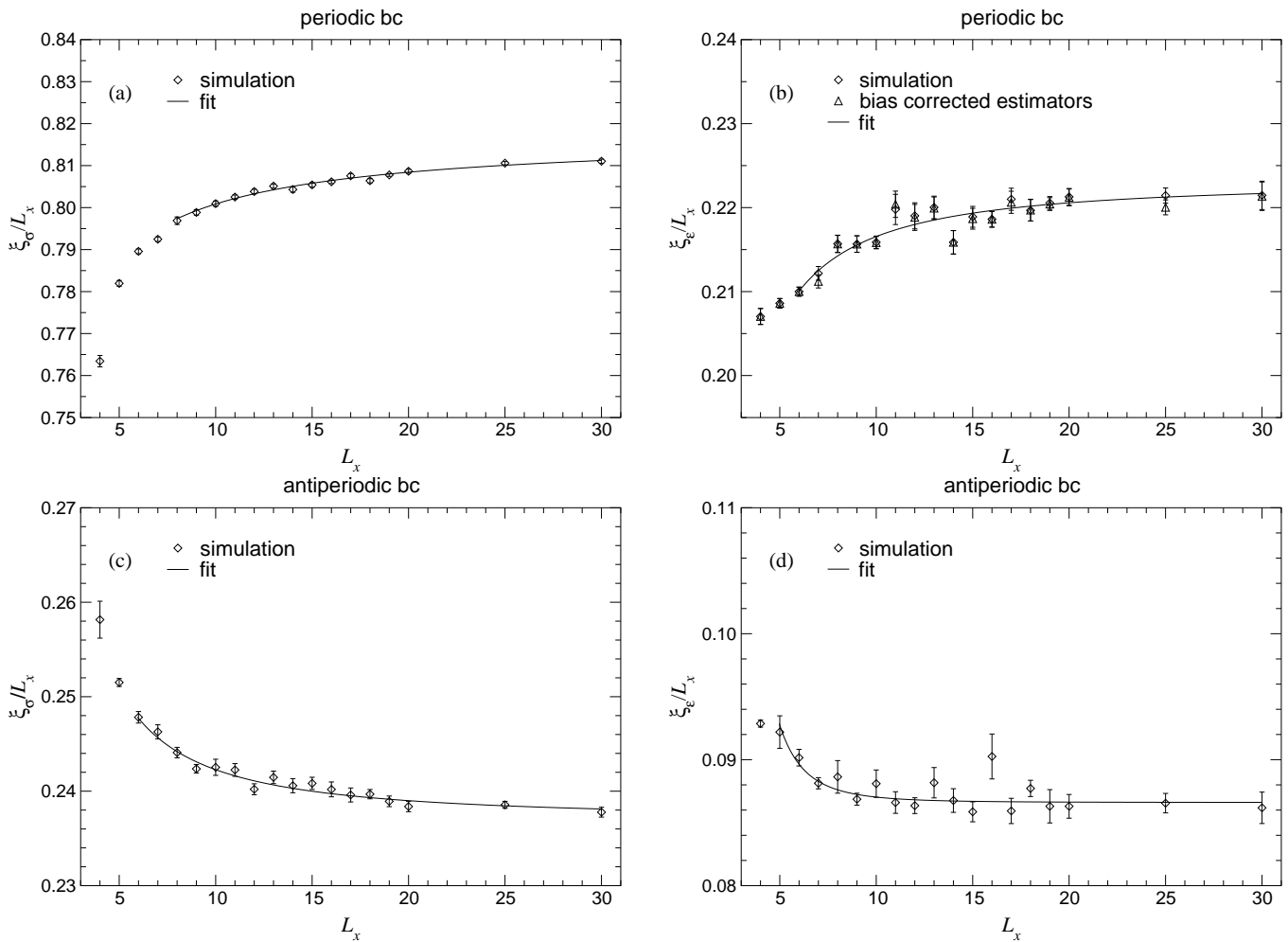

  \includegraphics[angle=-90,clip=true,width=\columnwidth,keepaspectratio=true]{xs3Dispd.eps}
  \hspace{\sepmod}
  \includegraphics[angle=-90,clip=true,width=\columnwidth,keepaspectratio=true]{xe3Dispd.eps}
  \includegraphics[angle=-90,clip=true,width=\columnwidth,keepaspectratio=true]{xs3Disad.eps}
  \hspace{\sepmod}
  \includegraphics[angle=-90,clip=true,width=\columnwidth,keepaspectratio=true]{xe3Disad.eps}
  \caption{Scaling of the amplitudes $\bar{\xi}_{\sigma/\epsilon}/L_x$
    for the Ising model. The solid lines show fits to the function Eq.\ 
    (\ref{fitform}); (a) and (b) show correlation lengths for the systems with
    periodic boundary conditions, (c) and (d) for the case of an antiperiodic
    boundary; (b) additionally contains bias corrected estimates according to Eq.\ 
    (\ref{xitild}).\label{is_amp}}
\end{figure*}

\paragraph*{Ising Model --} Simulations of the Ising model were done at an inverse
temperature given by a high-precision MC estimate of the bulk critical coupling in
three dimensions\cite{talapov:96}, $\beta_c=0.2216544(3)$. We use a temperature
reweighting technique to check for the influence of the uncertainty of $\beta_c$ on
the final results \cite{ferrenberg:88a,ferrenberg:88ae}.  We find it completely
negligible compared to the statistical errors for the case of the Ising model.  To
enable a proper FSS analysis including sub-leading terms we performed simulations for
transverse system sizes $L_x=4,\,5,\,\ldots,\,20,\,25,$ and $30$, scaling $L_z$
accordingly. Adapting the frequency of measurements to the autocorrelation times,
about $2\times10^6$ and $8\times10^6$ nearly independent measurements were recorded
for the systems with periodic and with antiperiodic boundary conditions,
respectively. Collecting the final estimates $\bar{\xi}$ for the correlation lengths
one ends up with a scaling plot like that shown in Fig.\ \ref{scale_ex}. The scaling
behavior is quite linear, however, as plots of the amplitudes $\bar{\xi}/L_x$ reveal,
corrections to the purely linear scaling behavior are clearly resolvable, cp.\ Fig.\ 
\ref{is_amp}.  As an aside, Fig.\ \ref{is_amp}(b) additionally shows jackknife bias
corrected estimators according to Eq.\ (\ref{xitild}); for the given length of time
series bias effects of our estimator Eq.\ (\ref{diffmethoddelta}) can clearly be
neglected.

\begin{table}[tb]
  \caption{Literature estimates for the inverse critical temperature $\beta_c$ and
    the critical exponents $\nu$ and $\gamma$ of the
    three-dimensional XY ($n=2$) and Heisenberg ($n=3$) models.\label{xytab}} 
 \begin{center}
   \begin{tabular*}{\linewidth}{@{\extracolsep{\fill}}lllll} \toprule
     $n$ & Method  & \multicolumn{1}{c}{$\beta_c$} & \multicolumn{1}{c}{$\nu$} &
     \multicolumn{1}{c}{$\gamma$} \\ \colrule
      & $\epsilon$-expansion \cite{guida:98a} & \multicolumn{1}{c}{---} & 0.6680(35) & 1.3110(70) \\
      & series \cite{butera:97a} & 0.45419(3) & 0.677(3) & 1.327(4) \\
      & series \cite{pelissetto:00a} & \multicolumn{1}{c}{---} & 0.67166(55) & 1.3179(11) \\
      & series \cite{butera:93a} & 0.45406(5) & \multicolumn{1}{c}{---} & \multicolumn{1}{c}{---} \\
      & series \cite{butera:95a} & 0.45420(6) & \multicolumn{1}{c}{---} & \multicolumn{1}{c}{---} \\
      \raisebox{2ex}[2ex]{$2$} & MC \cite{wj:90a} & 0.4542(1) & 0.670(2) & 1.319(2)  \\
      & MC \cite{ballesteros:96a} & 0.454165(4)  & 0.672(1) & 1.316(3)  \\     
      & MC \cite{hasenbusch:99a} & \multicolumn{1}{c}{---} & 0.6723(11) & 1.3190(22) \\
      & MC \cite{hasenbusch:90a} & 0.45421(8)   & \multicolumn{1}{c}{---} & \multicolumn{1}{c}{---} \\
      & MC \cite{gottlob:93a} & 0.454170(7)  & \multicolumn{1}{c}{---} & \multicolumn{1}{c}{---} \\ \colrule
      & weighted mean & 0.454167(3) & 0.67179(42) & 1.31839(82) \\ \colrule
      & $\epsilon$-expansion \cite{guida:98a} & \multicolumn{1}{c}{---} & 0.7045(55) & 1.3820(90) \\
      & series \cite{adler:93a} & 0.6929(1)    & 0.712(10)  & 1.400(10) \\
      & series \cite{butera:97a} & 0.69305(4)   & 0.716(2)   & 1.406(3) \\
     $3$ & MC \cite{peczak:91a} & 0.6929(1)    & 0.706(9)   & 1.390(23)  \\
      & MC \cite{chen:93a} & 0.693035(37) & 0.7036(23) & 1.3896(70)  \\
      & MC \cite{holm:93a,holm:93b} & 0.6930(1)    & 0.704(6)   & 1.389(14)  \\ 
      & MC \cite{ballesteros:96a} & 0.693002(12) & 0.7128(14) & 1.399(2) \\ \colrule
      & weighted mean & 0.69301(1) & 0.71129(98) & 1.3998(16) \\ \botrule
   \end{tabular*}
 \end{center}
\end{table}

Returning to the two-dimensional case for a moment, it is easy to see the source for
the leading correction term in the correlation length scaling. In the framework of
conformal field theory the effect of lattice discretization as well as the influence
of non-linearity of scaling fields that increase with the distance from criticality
(i.e.\ the thermodynamic limit in our case) can be included in considerations using
conformal perturbation theory \cite{henkel:book}. A formal perturbation expression
for the spin-spin correlation length including the effect of a perturbing operator
coupled with strength $a_k$ is to first order given by
\begin{equation}
  {\xi_\sigma}^{-1}=\frac{2\pi}{L}\left[x+2\pi a_k({\bf C}_{1k1}-{\bf C}_{0k0})
    {\left(\frac{2\pi}{L}\right)}^{x_k-2}\right],
  \label{conf_pert}
\end{equation} 
where the perturbing operator has dimension $x_k$ and the coefficients ${\bf
  C}_{nkn}$ result from the operator product expansion (OPE). One finds
\cite{reinicke:87a} that to lowest order the only non-vanishing amplitude belongs to
an operator that corresponds to the breaking of rotational symmetry by the square
lattice as compared to the continuum solution. It has dimension $x_k=4$ leading to
$1/L^2$ corrections, in agreement with the first-order expansion of the exact result
\cite{nightingale:76a}:
\begin{equation}
  {\xi_\sigma}^{-1}(L)=\frac{2\pi}{L}\left[\frac{1}{8}-2\pi\frac{1}{768\pi}
    {\left(\frac{2\pi}{L}\right)}^2\right].
\end{equation}

A similar effect will be present in the three-dimensional systems, but the correction
exponent can no longer be evaluated analytically. Fig.\ \ref{is_amp} shows that the
{\em sign\/} of the leading correction term is unchanged in three dimensions for the
systems with periodic boundary conditions, whereas it is reversed for the systems
with antiperiodic boundary. This stays true for the other O($n$) spin models
discussed below. To account for corrections to scaling we fit the correlation lengths
data to the functional form
\begin{equation}
  \xi(L_x)=AL_x+BL_x^{\kappa},
  \label{fitform}
\end{equation} 
treating the correction exponent $\kappa$ as an additional fit parameter. Due to the
presence of higher-order corrections, however, the resulting values of $\kappa$ have
to be taken as effective exponents, that will in general differ from Wegner's
correction exponent $\omega$. Therefore we decided to keep $\kappa$ as a parameter,
despite of existing field-theory estimates for $\omega$, cp.\ \cite{zinn-justin}. We
take into account the effect of neglecting higher-order correction terms by
successively dropping points from the small $L_x$ end while monitoring the
quality-of-fit parameters $\chi^2/g$ or $Q$ to find a compromise between fit
stability and precision of the final amplitudes $A$. The range of sizes $L_x$ used is
indicated by the range of the solid lines in Fig.\ \ref{is_amp}. Our results for the
scaling amplitudes and their ratios as listed in Table \ref{ratiotab} and the ratio
of scaling dimensions according to Eq.\ (\ref{x_hyper}) show precise agreement in the
sense of Eq.\ (\ref{ratio}) for the case of antiperiodic boundary conditions and
clear mismatch for a periodic boundary. This is in agreement with the results of
Henkel \cite{henkel:87a} and Weston \cite{weston:90}, but at a level of accuracy that
makes a casual coincidence very unlikely.

\begin{figure}[tb]
  \begin{center}
    \includegraphics[angle=-90,clip=true,width=\linewidth,keepaspectratio=true]{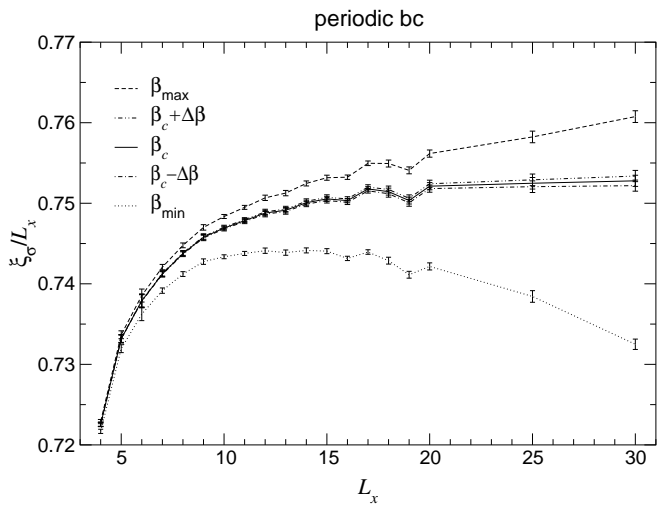}
  \end{center}
  \caption{Amplitude scaling of the spin-spin correlation length of the XY model with
    periodic boundary conditions. The spread curves show results of temperature
    reweighting for $\beta_c-\Delta\beta=0.454164$, $\beta_c+\Delta\beta=0.454170$,
    $\beta_\mathrm{min}=0.45406$ and $\beta_\mathrm{max}=0.45421$.\label{xy_beta_plot}}
\end{figure}

\paragraph*{XY Model --} The XY model is, as well as the Heisenberg models,
accessible to cluster update methods using the embedded cluster representation
\cite{wj:chem}, which we made use of. The simulations were performed at the coupling
$\beta_c=0.454167(3)$, which is an average of recent literature estimates, cp.\ Table
\ref{xytab}. Using the same transverse system sizes $L_x=L_y$ as for the Ising model,
but adjusting the lengths $L_z$ according to the different correlation length
amplitudes, we took between $1\times10^6$ and $16\times10^6$ measurements, using
measurement frequencies around $1/\tau_\mathrm{int}$ as above. Fig.\ 
\ref{xy_beta_plot} shows the amplitude scaling plot of the spin-spin correlation
length for periodic boundary conditions. The additional curves are results of a
temperature reweighting analysis, trying to judge the effect of critical coupling
uncertainties. The precision of the data is well illustrated by the fact that,
reweighting our results to the minimum and maximum estimated critical couplings,
respectively, cited in Table \ref{xytab}, results in a variation of the scaling
curves far beyond the range covered by the remaining statistical errors.
Nevertheless, reweighting to the $1\sigma$-range inverse temperatures
$\beta_c-\Delta\beta$ and $\beta_c+\Delta\beta$ as given above triggers deviations at
most comparable to the error estimates of the statistical analysis. The intermediate
maximum of the curve for $\beta_\mathrm{min}$, however, might be an artefact
indicating that $\beta_\mathrm{min}$ is already too far away from the simulation
temperature to allow for reliable reweighting. The effect of temperature variation is
generally observed to be smaller for the antiperiodic boundary systems; furthermore,
it is more important for the case of the spin-spin correlation length since here
statistical errors are clearly smaller than for the energy-energy correlation length
estimates. Thus, Fig.\ \ref{xy_beta_plot} shows the largest effect observed.

Fitting the final correlation length results $\bar{\xi}_{\sigma/\epsilon}$ to the
functional form Eq.\ (\ref{fitform}), we arrive at the final estimates for the
leading amplitudes given in Table \ref{ratiotab}. Comparing these to the ratio of
scaling dimensions resulting from the averaged critical exponent estimates of Table
\ref{xytab} and Eq.\ (\ref{x_hyper}), we again find Eq.\ (\ref{ratio}) confirmed for
antiperiodic boundary conditions only; this behavior is obviously not specific to the
Ising model.

\paragraph*{Heisenberg Model --} The $n=3$ Heisenberg model case is treated analogously
to the XY model. Table \ref{xytab} gives the critical parameter estimates used for
the simulations and comparison. With statistics similar to that for the $n=1$ and
$n=2$ cases, the simulations confirm the findings of the Ising and XY models, cp.\ 
Table \ref{ratiotab} for details. For the case of the energy-energy correlation
length of the systems with periodic boundary conditions the gathered statistics did
not suffice for a stable non-linear fit including corrections according to Eq.\ 
(\ref{fitform}). We thus performed a simple linear fit dropping the correction term.
This, however, results in an error estimate which is not quite realistic and,
furthermore, induces a systematic underestimation of the amplitude since one expects
$B_\epsilon<0$, cp.\ Fig. \ref{is_amp}(b). From the results of the other models this
effect is estimated to be about $2\sigma$-$3\sigma$ in magnitude.

\paragraph*{O(10) Model --} To gain additional evidence and in order to facilitate
considerations concerning the $n\rightarrow\infty$ limit, giving a clear picture of
systematic dependencies on the parameter $n$, we also simulated the $n=10$
generalized Heisenberg model. Since, of course, in the past much less effort has gone
into the investigation of the O($n$) model with $n>4$, there are quite few estimates
of the critical coupling. We thus here use a single high-temperature series estimate
of $\beta_c=2.42792(8)$ \cite{butera:97a}. The implementation of the Wolff cluster
update algorithm has to cope with the technical intricacy of generating pseudo-random
numbers equally distributed on a hyper-sphere, see Appendix A for details. Due to
this complication we only simulated systems up to a transversal size of $L_x=20$ and
reduced the number of measurements to $2\times10^6$. The critical exponents for
comparison, given by a plain average over some recent estimates
\cite{sokolov:95a,butera:97a,kleinert:99a}, are:
\begin{equation}
  \nu=0.8713(75),\;\gamma=1.721(14).
  \label{o10nu}
\end{equation}
Table \ref{ratiotab} shows again agreement between amplitude and exponent ratios only
for the case of antiperiodic boundaries. Note that, as critical exponent estimates
become rare with increasing $n$, the correlation length ratio estimate already
reaches the precision of the scaling dimension ratio estimate. Checking the influence
of the critical coupling uncertainty we find it only important compared to
statistical errors in the case of the spin-spin correlation length for periodic
boundary systems; the results reweighted to $\beta_\pm=\beta_c\pm\Delta\beta$ are
$A_\sigma^-=0.670805(56)$ and $A_\sigma^+=0.671432(65)$, respectively. This, however,
does not noticeably influence the error of the ratio estimate, since here the error of
the estimate of $A_\epsilon$, which is much larger, is dominant.

We thus find the linear amplitude-exponent relation Eq.\ (\ref{ratio}) confirmed for
several spin models in three dimensions with the peculiarity that one has to insert a
seam of antiferromagnetic bonds along the $T^2$-directions to restore the
two-dimensional situation.

\begin{table}[bt]
  \caption{FSS amplitudes of the correlation lengths of O($n$) spin models
    on the $T^2\times\mathbb{R}$ geometry.\label{ratiotab}}
  \begin{center}
    \begin{tabular*}{\linewidth}{@{\extracolsep{\fill}}clll}
      \toprule
      Model & & \multicolumn{1}{c}{periodic bc} & \multicolumn{1}{c}{antiperiodic bc} \\ \colrule
      & $A_\sigma$    & 0.8183(32) & 0.23694(80) \\
      & $A_\epsilon$  & 0.2232(16)  & 0.08661(31) \\
      \raisebox{1ex}[-1ex]{Ising} & $A_\sigma/A_\epsilon$ & 3.666(30) & 2.736(13) \\ 
      & $x_\epsilon/x_\sigma$ &  \multicolumn{2}{c}{2.7264(13)} \\ \colrule
      & $A_\sigma$    & 0.75409(59) & 0.24113(57) \\
      & $A_\epsilon$  & 0.1899(15)  & 0.0823(13) \\ 
      \raisebox{1ex}[-1ex]{XY} & $A_\sigma/A_\epsilon$ & 3.971(32) & 2.930(47) \\ 
      & $x_\epsilon/x_\sigma$ & \multicolumn{2}{c}{2.9136(38)} \\ \colrule
      & $A_\sigma$    & 0.72068(34) & 0.24462(51) \\
      & $A_\epsilon$  & 0.16966(36)  & 0.0793(20) \\
      \raisebox{1ex}[-1ex]{Heisenberg} & $A_\sigma/A_\epsilon$ & 4.2478(92) & 3.085(78) \\ 
      & $x_\epsilon/x_\sigma$ &  \multicolumn{2}{c}{3.0891(79)} \\ \colrule
      & $A_\sigma$    & 0.671107(59) & 0.25865(46) \\
      & $A_\epsilon$  & 0.1350(23)  & 0.07096(107) \\
      \raisebox{1ex}[-1ex]{$n=10$} & $A_\sigma/A_\epsilon$ & 4.971(83) & 3.645(55) \\ 
      & $x_\epsilon/x_\sigma$ &  \multicolumn{2}{c}{3.615(70)} \\
      \botrule
    \end{tabular*}
  \end{center}
\end{table}

\section{Results: ``Meta'' Amplitudes}

Comparing our results for the three-dimensional geometry $S^1\times
S^1\times\mathbb{R}$ to the CFT conjecture for the case of two dimensions, we are
interested in the respective ranges of validity in terms of the classification of
universality aspects given above in the Introduction. The fact that our simulations
of the isotropic lattice representation of the O($n$) universality classes give
results in agreement with the strongly anisotropic quantum Hamiltonian representation
used by Henkel in his transfer matrix calculations for the case of the Ising model
\cite{henkel:86a,henkel:87a,henkel:87b}, indicates that the considered amplitude {\em
  ratios\/} are universal, i.e.\ (i') holds. Apart from that, Henkel
\cite{henkel:86a} explicitly checked for universality of amplitude ratios by the
insertion of an irrelevant perturbing operator and found it confirmed for both cases
of boundary conditions. However, strictly speaking, there is no proof of universality
for the cases $n>1$.  The universality aspect (i) above, i.e.\ universality of the
amplitudes themselves, could not be checked in Henkel's calculations, because the
quantum Hamiltonian is only defined up to an overall normalization constant.
Yurishchev \cite{yuri:94a,yuri:97a} considered the behavior of an anisotropic Ising
model and found varying correlation lengths amplitudes on variation of the ratios of
couplings in the different directions. This, however, is no argument against
amplitude universality since anisotropy is represented by marginal instead of
irrelevant operators. On the other hand, amplitude {\em ratios\/} stay universal even
with respect to those marginal perturbations, in consistency with Henkel's strongly
anisotropic Hamiltonian limit calculations. In fact it has been argued that for all
systems below their upper critical dimension correlation length scaling amplitudes
are universal quantities \cite{privman:84a}.

Having found very good agreement in three dimensions between ratios of correlation
lengths and scaling dimensions according to Eq.\ (\ref{ratio}) for the case of
antiperiodic boundary conditions, it is interesting to see what the overall,
operator-independent, ``meta'' amplitude ${\cal A}$ according to:
\begin{equation}
  \xi_{\sigma/\epsilon}=A_{\sigma/\epsilon}L_x=
  \frac{{\cal A}}{x_{\sigma/\epsilon}}L_x,
\end{equation}
that was ${\cal A}=1/2\pi$ for two-dimensional periodic systems, cp.\ Eq.\ 
(\ref{xi_par}), becomes in three dimensions, in particular whether it is again model
independent. Since our results for the spin-spin correlation lengths are always more
precise than those for energy-energy correlation lengths, we use $\bar{\xi}_\sigma$
to determine ${\cal A}$. The estimates for the spin-spin scaling dimension $x_\sigma$
resulting from the corresponding estimates of bulk critical exponents $\nu$ and
$\gamma$ listed in Tables \ref{is_exp_tab} and \ref{xytab} and Eq.\ (\ref{o10nu})
are $x_\sigma=0.5182(4)$ (Ising), $x_\sigma=0.5188(9)$ (XY), $x_\sigma=0.5160(17)$
(Heisenberg), and $x_\sigma=0.512(12)$ ($n=10$), respectively. Thus, inserting our
results for $A_\sigma$ listed in Table \ref{ratiotab}, we obtain for the ``meta''
amplitudes ${\cal A}(n)$:
\begin{equation}
  {\cal A}=A_\sigma x_\sigma=\left\{
    \begin{array}{l@{\hspace{0.5cm}}l}
      0.12278(43) & \mbox{Ising} \\
      0.12510(37) & \mbox{XY} \\
      0.12622(49) & \mbox{Heisenberg} \\
      0.1325(30) & \mbox{$n=10$} \\
    \end{array}
  \right. .
  \label{metaamp}
\end{equation}

\begin{figure}[tb]
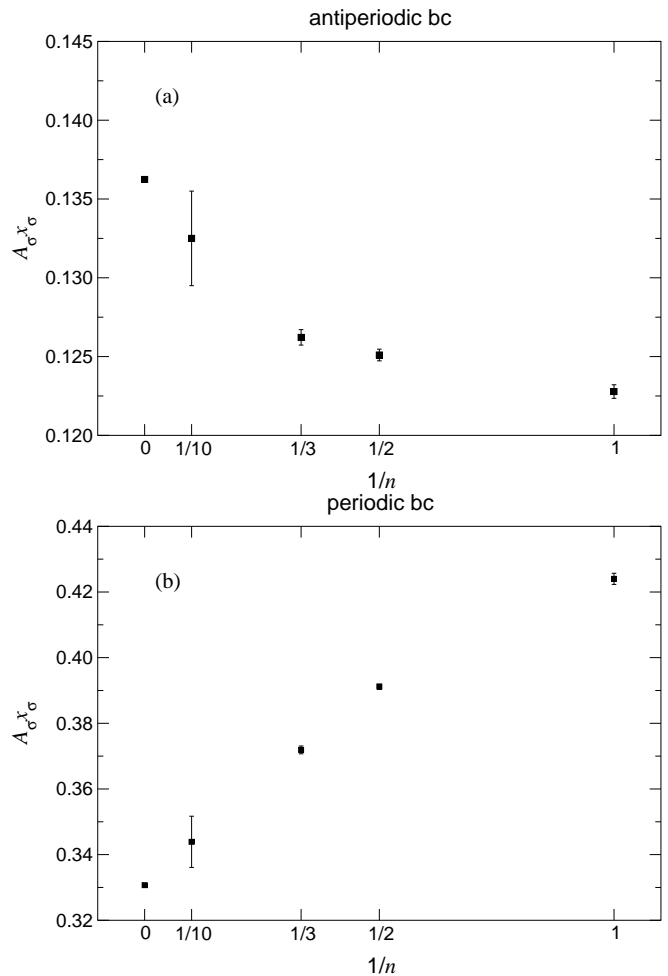

  \begin{center}
    \includegraphics[angle=-90,clip=true,width=\linewidth,keepaspectratio=true]{amplit.eps}
    \includegraphics[angle=-90,clip=true,width=\linewidth,keepaspectratio=true]{amplit2.eps}
  \end{center}
  \caption{(a) ``Meta'' amplitudes ${\cal A}$ for antiperiodic boundary conditions
    according to Eq.\ (\ref{metaamp}) as a function of the order parameter dimension
    $n$; (b) The same combination $A_\sigma x_\sigma$ for periodic boundary
    conditions according to Eq.\ (\ref{metaamp2}).\label{amp_fig}}
\end{figure}

These values can additionally be compared with an analytical result that is available
for the case of the spherical model, which is commonly believed to be identical to
the $n\rightarrow\infty$ limit of the O($n$) spin model \cite{stanley:68a}. Again
using the Hamiltonian formulation, Henkel and Weston \cite{henkel:88a,henkel:92a}
found that the amplitude exponent relation Eq.\ (\ref{ratio}) is valid for the
spherical model on $S^1\times S^1\times\mathbb{R}$ for {\em both\/} kinds of boundary
conditions, periodic and antiperiodic. This is due to the fact that the quantum
Hamiltonian factorizes into a set of uncoupled harmonic oscillators. The amplitude
${\cal A}$ for the case of antiperiodic boundary conditions was found to be ${\cal
  A}\approx 0.13624$ \cite{henkel:92a,allen:93a}. Plotting this value together with
the finite-$n$ results of Eq.\ (\ref{metaamp}) shows an apparently smooth variation
of the ``meta'' amplitudes with the order parameter dimension $n$, the eyeball
extrapolation of the finite-$n$ values to $1/n\rightarrow 0$ matching the spherical
model result, cp.\ Fig.\ \ref{amp_fig}(a).  Facing this variation, the hypothesis of
a ``hyperuniversal'' amplitude ${\cal A}(n)={\cal A}$ that does not depend on $n$, as
was the case for the two-dimensional systems, can be clearly ruled out. Thus, type
(iii) universality of the classification above gets broken when passing from two to
three dimensions. The matching of the finite-$n$ values with the universal spherical
model amplitude, on the other hand, indicates universality also of the finite-$n$
amplitudes and thus universality of type (i) above.

Even without a scaling law of the type Eq.\ (\ref{ratio}) being valid for the case of
periodic boundary conditions, one can nevertheless plot the corresponding combination
$A_\sigma x_\sigma$ for this case also, as is illustrated in
Fig.\ \ref{amp_fig}(b). The values are:
\begin{equation}
  A_\sigma x_\sigma=\left\{
    \begin{array}{l@{\hspace{0.5cm}}l}
      0.4240(17) & \mbox{Ising} \\
      0.3912(7)  & \mbox{XY} \\
      0.3719(12) & \mbox{Heisenberg} \\
      0.3439(78) & \mbox{$n=10$} \\
    \end{array}
  \right. .
  \label{metaamp2}
\end{equation}
The corresponding value for the spherical model is given by
$A_\sigma x_\sigma\approx 0.3307$, cp.\ \cite{brezin:82a,henkel:92a}. The finite-$n$
values again run smoothly into the spherical model limit.

\section{The limit of infinite spin dimensionality}

\begin{figure}[tb]
  \begin{center}
    \includegraphics[angle=-90,clip=true,width=\linewidth,keepaspectratio=true]{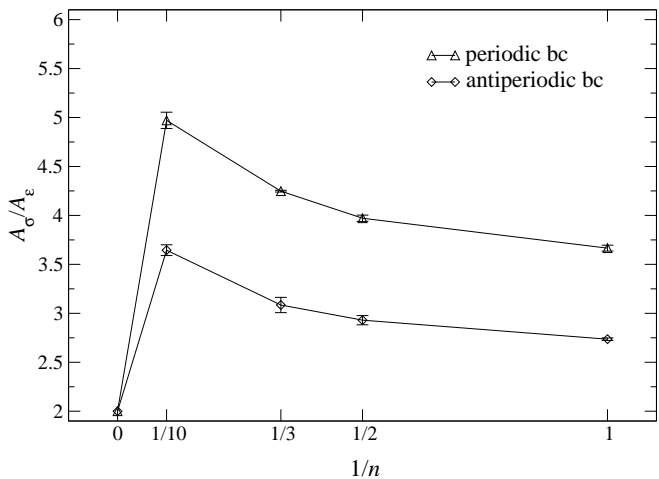}
  \end{center}
  \caption{Correlation lengths ratios as function of the order parameter dimension
    $n$ for periodic and antiperiodic boundary conditions.\label{ratio_fig}}
\end{figure}

While the finite-$n$ amplitudes of Fig.\ \ref{amp_fig} fit well to the spherical
model result, this is not the case for the correlation lengths ratios themselves.
From inspection of Fig.\ \ref{ratio_fig} the smooth variation of correlation length
ratios for finite $n$ does not fit at all to the spherical model result of Henkel and
Weston \cite{henkel:88a,henkel:92a} that gives a ratio $A_\sigma/A_\epsilon=2$ for
{\em both}, periodic and antiperiodic boundary conditions. By eyeball extrapolation
one would instead expect the amplitude ratios to reach values around $4$ for
antiperiodic and around $5\frac{1}{3}$ for periodic boundary conditions in the limit
$n\rightarrow\infty$.  And indeed, accepting the validity of a linear
amplitude-exponent relation according to Eq.\ (\ref{ratio}) for the case of
antiperiodic boundary conditions and using the usual relations for the connection
between scaling dimensions and bulk critical exponents, namely Eq.\ 
(\ref{xdurchexp}), one would expect $x_\sigma=1/2$ and $x_\epsilon=2$ since
$\beta=1/2$, $\nu=1$ and $\alpha=-1$ for the spherical model.  The resulting ratio
$x_\epsilon/x_\sigma=4$ perfectly agrees with the eyeball extrapolation of our
finite-$n$ data. However, by inspection of the energy-energy correlation function in
the Hamiltonian limit and using factorization arguments, Henkel \cite{henkel:88a}
conjectured $x_\epsilon=1$ instead, resulting in the ratio $A_\sigma/A_\epsilon=2$,
in contrast to the relation Eq.\ (\ref{xdurchexp}). Taking into account the obvious
agreement of eyeball extrapolation and spherical model calculation for the amplitudes
${\cal A}(n)$ that were calculated from the spin-spin correlation length amplitude as
${\cal A}(n)=A_\sigma x_\sigma$, cp.\ Fig.\ \ref{amp_fig}, it becomes obvious that
the mismatch is entirely due to the behavior of the energy-energy correlations. Note
also that, since the specific heat is constant in the low-temperature phase of the
spherical model in three dimensions, interpreting this as an effectively vanishing
specific-heat exponent $\alpha'=0$ leads to an effective energetic scaling dimension
$x_\epsilon'=1$. This, in fact, implies a violation of the scaling relation Eq.\ 
(\ref{xdurchexp}), which is of the hyper-scaling type, for the case of the spherical
model.

Puzzled by this striking mismatch, we performed a roughly explorative MC simulation
directly in the spherical model, which rendered results in qualitative agreement with
an amplitude ratio of $A_\sigma/A_\epsilon=2$ as suggested by the analytical
calculation. Then, it is natural to ask whether there is a contradiction with
Stanley's result on the equivalence of the $n\rightarrow\infty$ limit of the O($n$)
model and the spherical model \cite{berlin:52}, which has been, after some debate
over mathematical subtleties \cite{helfand:69a}, rigorously proven \cite{kac:71a}.
The precise statement that can be proven is the identity of the partition functions
or, equivalently, free energies of the two models in the thermodynamic limit for the
whole temperature range, even independent of the order of taking the limits
$n\rightarrow\infty$ and $N\rightarrow\infty$ (the thermodynamic limit). Since
multi-point correlation functions do not follow from the (source-free) partition
function, this does not say anything about the behavior of these functions in those
two models. A direct calculation in the spherical model, cf.\ Appendix B, results in
a simple factorization property of the long distance behavior of the connected
energy-energy correlation function for all temperatures in one and two dimensions and
in the high-temperature phase down to $T_c$ in three dimensions. If the four-point
function of the spherical-model spins is denoted by $C_{ijkl}$, one has:
\begin{eqnarray}
  & C_{i\,i+1\;j\,j+1}-C_{i\,i+1}^2 & = C_{i\,j}C_{i+1\,j+1}+C_{i\,j+1}C_{i+1\,j} \nonumber \\
  & & \longrightarrow 2C_{i\,j}^2,\; |j-i|\rightarrow\infty,
\end{eqnarray}
where $C_{ij}$ are the corresponding two-point functions. This confirms Henkel's
results for the Hamiltonian formulation \cite{henkel:88a} on more general grounds.

Considering the $n\rightarrow\infty$ limit of the O($n$) model, on the other hand,
reveals that the connected part of the energy-energy correlation function {\em
  vanishes\/} in the first-order saddle-point approximation that is being used for
the comparison of the two models, cf.\ Appendix C. This is in agreement with general
considerations for the large $n$ model by Br\'ezin \cite{brezin:82a}. For the case of
the one-dimensional spin chain, the connected energy-energy correlation function even
vanishes exactly for all $n$, so that one can rule out an agreement of the two limits
to higher order of the steepest-descent expansion in this case. Thus the mismatch of
finite-$n$ extrapolations and spherical model results of Fig.\ \ref{ratio_fig} has
some well-defined mathematical reason.

\begin{figure}[tb]
  \begin{center}
    \includegraphics[angle=-90,clip=true,width=\linewidth,keepaspectratio=true]{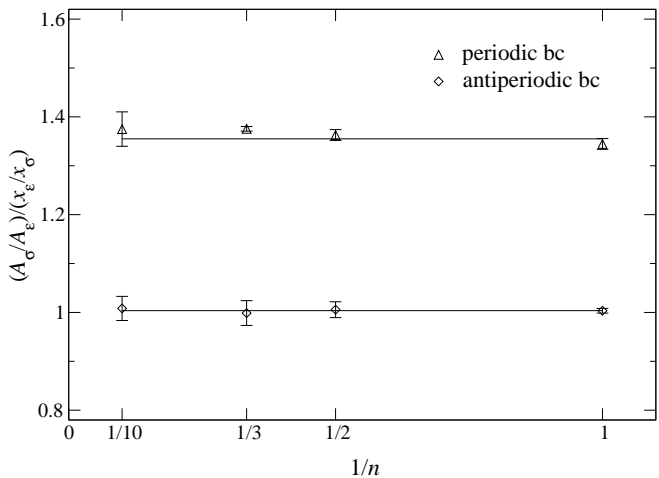}
  \end{center}
  \caption{Matching of correlation lengths ratios $A_\sigma/A_\epsilon$ and inverse
    scaling dimension ratios $x_\epsilon/x_\sigma$ for the two kinds of boundary
    conditions as a function of the order parameter dimension $n$. The horizontal
    lines show fits to a constant as discussed in the text.\label{ratio_ratio_fig}}
\end{figure}

Starting from the observation that the curves of Fig.\ \ref{ratio_fig} for the
amplitude ratios seem to be quite parallel as a function of (finite) $n$ for the both
kinds of boundary conditions, we also plotted the collapsed ratio
$\frac{A_\sigma/A_\epsilon}{x_\epsilon/x_\sigma}$ that should be unity if the
amplitude-exponent relation Eq.\ (\ref{ratio}) holds true.  Inspecting Fig.\ 
\ref{ratio_ratio_fig}, this is, according to our above results, of course the case
for antiperiodic boundary conditions. Moreover, and a priori somewhat unexpected,
this ratio seems to be also quite constant for the case of a periodic boundary,
stabilizing around a value compatible with $4/3$ within statistical errors. Note that
the exceptionally small error of the value for $n=3$ (the Heisenberg model) and its
apparent deviation towards a larger ratio is due to the impossibility to fit the
$n=3$ energy-energy correlation lengths to a scaling law including a correction term
as mentioned in Sec.\ IV. Statistically, the data are consistent with a fit to a
constant ($Q=0.08$), and perfectly so when dropping the $n=3$ point ($Q=0.4$).

In view of this observation one might argue that the asymptotic scaling relation Eq.\ 
(\ref{ratio}) in three dimensions has to be replaced by a generalized ansatz of the
form
\begin{equation}
  \frac{\xi_\sigma}{\xi_\epsilon}=R\frac{x_\epsilon}{x_\sigma},
  \label{ratio_mod}
\end{equation}
with an overall, model independent factor $R$ that depends only on the boundary
conditions and happens to be just $1$ for the case of an antiperiodic boundary. For
the amplitude scaling law this would lead to an asymptotic form
\begin{equation}
  \xi_{\sigma/\epsilon}(n)=R\frac{{\cal A}(n)}{x_{\sigma/\epsilon}}L_x,
\end{equation}
cp.\ Eq.\ (\ref{xi_par}). Accepting such a generalized ansatz, a least-squares fit of
the collapsed ratios of Fig.\ \ref{ratio_ratio_fig} to a constant $R$ gives
$R=1.0037(45)$ for antiperiodic boundary conditions, underlining the validity of the
original amplitude-exponent relation Eq.\ (\ref{ratio}), or alternatively Eq.\ 
(\ref{ratio_mod}) with $R=1$, for this case. For the periodic-boundary systems, on
the other hand, we arrive at $R=1.3546(76)$ (omitting the $n=3$ point), indeed
statistically consistent with the conjectured value of $4/3$.

This somewhat diminishes the at
first sight apparently exceptional importance of choosing antiperiodic boundary
conditions in three dimensions. Taking into account the smooth amplitude variation of
Fig.\ \ref{amp_fig}(b) the same universality statements hold for periodic and for
antiperiodic boundary conditions.

\section{Conclusions}

We performed extensive MC simulations for several representatives of the class of
O($n$) spin models. Concentrating on the geometry of three-dimensional slabs
$S^1\times S^1\times\mathbb{R}$ we found a simple inversely linear relation between
the leading scaling amplitudes of the correlation lengths of the magnetization and
energy densities and the corresponding scaling dimensions valid to high accuracy for
the Ising ($n=1$), XY ($n=2$), Heisenberg ($n=3$), and $n=10$ generalized Heisenberg
models for {\em antiperiodic\/} boundary conditions along the torus directions. This
is the analogue of the CFT result in two dimensions with periodic boundary conditions
applied. There is evidence for the universality not only of amplitude ratios (type
(i') of our classification in the Introduction), but also of scaling amplitudes
themselves (type (i)). To definitely decide the question whether universality in the
sense (ii) above, i.e.\ condensation of all operator dependent information in the
scaling dimensions, is present, further operators would have to be considered.
Independence, apart from changes in the scaling dimension, of the scaling amplitudes
from the model under consideration, i.e.\ type (iii) universality, is explicitly
broken for three dimensions as compared to the two-dimensional case: we find a smooth
variation of the overall ``meta'' amplitudes ${\cal A}(n)=A_\sigma(n)x_\sigma(n)$,
depending on the order-parameter dimension $n$. It might be interesting to consider
further classes of models, such as for example Potts models, to see whether any of
these properties are specific to the O($n$) spin model class.

Considering the deviation of the periodic boundary correlation lengths ratios from
the corresponding inverse scaling dimension ratios, the validity of a scaling law of
the form Eq.\ (\ref{ratio}) can be definitely ruled out for this case. Generalizing
this ansatz with an overall factor $R$ depending on boundary conditions as in Eq.\ 
(\ref{ratio_mod}), however, we find it fulfilled also for the case of periodic
boundaries with a factor $R$ independent from $n$ and taking a value compatible with
$4/3$.  In view of that, the fact that $R=1$ for the case of antiperiodic boundary
conditions might be rather a coincidence than a ``deep'' physical property. Taking
into account that in two dimensions the corresponding prefactors are specific to the
operators considered, cp.\ Eq.\ (\ref{2Danti}), makes it probable that a similar
behavior occurs in three dimensions, destroying type (ii) universality. It might be
interesting to analyze the behavior of correlation lengths in the four-dimensional
geometry $S^1\times S^1\times S^1\times\mathbb{R}$ to check whether a scaling law of
the generalized form Eq.\ (\ref{ratio_mod}) can be retained and if so, how the factor
$R$ depends on the dimensionality of the lattice.

Trying to match our finite $n$ results with analytical calculations for the spherical
model we found a striking mismatch of the data concerning energy-energy correlations.
Inspecting the four-point functions directly in the spherical model and the
O($n\rightarrow\infty$) model limit we find that both results do not match to first
order of the saddle-point approximation in general dimensions and to all orders in
one dimension. Thus, the idea of equivalence of the two models has to be limited to
its original extent, namely the identity of partition functions in the thermodynamic
limit. Quantities not directly related to the partition function, like multi-point
correlation functions, do not necessarily have to coincide. Further work has to be
done to possibly evaluate exactly the correlation lengths ratios in the
$n\rightarrow\infty$ limit for both sorts of boundary conditions.

Since, still, there is no explanation of the findings concerning the correlation
lengths ratios for finite $n$ in terms of a field-theoretic or otherwise exact
approach, we would like to encourage further research in this direction.

\begin{acknowledgments}
  The authors thank K.\ Binder, J.\ L.\ Cardy, and M.\ Henkel for useful discussions.
  MW gratefully acknowledges support by the ``Deutsche Forschungsgemeinschaft''
  through the ``Graduiertenkolleg Quantenfeldtheorie''.
\end{acknowledgments}

\appendix

\section{Equal distribution of random numbers on a hyper-sphere}

Consider a probability density in polar co-ordinates $f(\phi,\theta)$ equally
distributed on the $2$-sphere $S^2$, i.e.:
\begin{equation}
  \frac{f(\phi,\theta)\;\mathrm{d}\phi\,\mathrm{d}\theta}
  {\sin\theta\;\mathrm{d}\phi\,\mathrm{d}\theta}=\mathrm{const}.
\end{equation} 
Factorizing $f(\phi,\theta)=p(\phi)\,q(\theta)=\mathrm{const}\cdot\sin\theta$, and
taking into account the normalization condition $\int\mathrm{d}\Omega
f(\phi,\theta)=1$, one finds:
\begin{equation}
  f(\phi,\theta)=p(\phi)\,q(\theta)=\frac{1}{2\pi}\cdot\frac{1}{2}\sin\theta.
\end{equation}
Pseudo-random number generators usually generate numbers equally distributed in the
unit interval $[0,1]$. How does this transform to an arbitrary distribution? Let a
random variable $z$ be distributed with a density $g(z)$ and transform according to
$z'=\omega(z)$; the density $h(z')$ then follows from the equation
\begin{equation}
  g(z)\,\mathrm{d}z=h(z')\,\mathrm{d}z'=h(\omega(z))\,\omega'(z)\,\mathrm{d}z.
\end{equation}
Thus, for random numbers $z$ equally distributed in $[0,1]$ the transformation
$\theta=\arccos(1-2z)$ gives the desired distribution
$q(\theta)=\frac{1}{2}\sin\theta$. This form is being used for the simulations of the
$n=3$ Heisenberg model. For general polar co-ordinates in $\mathbb{R}^n$, $x_1 =
r\cos\theta_1$, $x_2 = r\sin\theta_1\cos\theta_2$, up to $x_n =
r\sin\theta_1\cdots\sin\theta_{n-1}$, where $0\le\theta_i\le\pi$,
$0\le\theta_{n-1}<2\pi$ is understood, the volume element is given by:
\begin{equation}
  \mathrm{d}V=r^{n-1}\sin^{n-2}\theta_1\,\sin^{n-3}\theta_2\cdots\sin\theta_{n-2}\;
  \mathrm{d}r\,\prod_i\mathrm{d}\theta_i,
\end{equation}
so that one has for the factors $f^{(i)}(\theta_i)$ of an equally distributed density
$f(\theta_1,\ldots,\theta_{n-1}) = \prod_i f^{(i)}(\theta_i)$:
\begin{eqnarray}
  & & f^{(i)}(\theta_i)  = 
  \frac{1}{\gamma(n-i-1)}\sin^{n-i-1}\theta_i,\;\;i<n-1, \nonumber\\
  & & f^{(n-1)}(\theta_{n-1})  =  \frac{1}{2\pi},
\end{eqnarray}
with normalization factors
$\gamma(k)=\sqrt{\pi}\,\Gamma(\frac{k+1}{2})/\Gamma(\frac{k}{2}+1)$. Thus, for $z_i$
equally distributed in $[0,1]$ the transformations $z_i(\theta_i)$ are given by:
\begin{equation}
  z_i(\theta_i)\equiv\mathrm{int}(\theta_i)=\frac{1}{\gamma(n-i-1)}\int\mathrm{d}\theta_i\,
  \sin^{n-i-1}\theta_i,
\end{equation}
for $i<n-1$. The integrals can be evaluated analytically for each $\theta_i$. There
is, however, no closed form expression for the {\em inverse\/} transformation
$\theta_i(z_i)$ that is needed to generate random vectors equally distributed on the
hyper-sphere $S^{n-1}$.  The trivial workaround solution of sampling equally
distributed in the hyper-cube $L^n=[-1,1]\times\cdots\times[-1,1]$, discarding the
complement $L^n\backslash B^n$ and projecting the remaining points on the sphere
$S^{n-1}$, suffers from asymptotically vanishing efficiency, since the ratio of used
to discarded volumes vanishes with increasing $n$ exponentially as
$\pi^{n/2}/2^n\Gamma(\frac{n}{2}+1)$.  We thus resorted to a numerical inversion of
$z_i(\theta_i)$ using interpolation between the pre-calculated points of a binary
tree.

\section{Energy-energy correlation function in the spherical model}

Consider the spherical model of Berlin and Kac \cite{berlin:52} consisting of
``spins'' $\epsilon_i\in\mathbb{R}$ with the constraint:
\begin{equation}
  \sum_{i=1}^N\epsilon_i^2=N,
  \label{constraint}
\end{equation}
where $N$ denotes the number of lattice sites. For ease of reference we use the
notation of the original paper here; thus, the $\epsilon_i$ are not to be confused
with the local energy densities defined above in Eq.\ (\ref{e_dens}). The Hamiltonian
is:
\begin{equation}
  {\cal H}=-J\sum_{\langle ij\rangle}\epsilon_i\epsilon_j.
\end{equation}
Using the Fourier representation of the $\delta$-constraint Eq.\ (\ref{constraint})
the partition function can be written as:
\begin{eqnarray}
  Z_N & = & \frac{A_N^{-1}}{2\pi i}\int_{\alpha_0-i\infty}^{\alpha_0+i\infty}{\rm
  d}s\;e^{Ns} {\int\cdots\int}{\rm d}\epsilon_1\cdots{\rm d}\epsilon_N \nonumber \\
  & & \times \exp(-s\sum_i\epsilon_i^2+K\sum_{\langle ij\rangle}\epsilon_i\epsilon_j),
  \label{sph_part}
\end{eqnarray}
choosing $\alpha_0$ such that the singularities in $s$ of the integrand are excluded
from the integration volume. $A_N$ ensures the correct normalization of the integral
measure and $K=\beta J$ denotes the coupling. Diagonalizing the quadratic form
$\sum_{\langle ij\rangle} \epsilon_i\epsilon_j$ with eigenvalues $\lambda_j$ via an
orthogonal transformation $\epsilon_i=\sum_j V_{ij}\,y_j$, the Gaussian integration
over the $\epsilon_i$ can be performed:
\begin{eqnarray}
  & \displaystyle{{\int\cdots\int}{\rm d}y_1\cdots{\rm
      d}y_N\,\exp[-\sum_j(s-K\lambda_j)y_j^2]}
  & \nonumber \\ & \displaystyle{=\pi^{N/2}\exp[-\frac{1}{2}\sum_j\ln(s-K\lambda_j)]} &
  \label{gaussian}
\end{eqnarray}
so that,
\begin{eqnarray}
  Z_N & = & A_N^{-1}\pi^{N/2}2Ke^{-\frac{1}{2}N\ln2 K}\frac{1}{2\pi
    i}\int_{z_0-i\infty}^{z_0+i\infty}{\rm d} z \nonumber \\
  & & \times \exp[N2Kz-\frac{1}{2}\sum_{j=1}^N\ln(z-\frac{1}{2}\lambda_j)],
  \label{spher_part}
\end{eqnarray}
where $s=2Kz$. This expression can be evaluated in the saddle point limit
$N\rightarrow\infty$ depending on the distribution of the eigenvalues $\lambda_i$ for
a given lattice. Now consider the two-point function,
\begin{equation}
  C_{ij}\equiv\langle\epsilon_i\epsilon_j\rangle=\sum_{r,s} V_{ir}V_{js}\langle y_r
  y_s\rangle=\sum_r V_{ir}V_{jr} \langle y_r^2\rangle,
  \label{spher_two}
\end{equation}
where the last equality follows from the symmetry of the partition function Eq.\ 
(\ref{sph_part}). Compared to the Gaussian integration Eq.\ (\ref{gaussian}) the
insertion of a factor $y_r^2$ in the integrand gives an additional factor of:
\begin{equation}
  \frac{1}{2(s-K\lambda_r)}=\frac{1}{4K(z-\frac{1}{2}\lambda_r)}.
  \label{pull_factor}
\end{equation}
The corresponding integral over $z$ can also be evaluated in the saddle point
approximation \cite{berlin:52}. Now, analogously, consider the four-point function:
\begin{equation}
  C_{ijkl}\equiv\langle\epsilon_i\epsilon_j\epsilon_k\epsilon_l\rangle=
  \sum_{r,s,t,u} V_{ir}V_{js}V_{kt}V_{lu}\langle y_r y_s y_t y_u\rangle.
\end{equation}
Here, again, only paired occurrences of the $y_m$ survive due to the inversion
symmetry:
\begin{eqnarray}
  & C_{ijkl} & = \sum_{r} V_{ir}V_{jr}V_{kr}V_{lr}\langle y_r^4\rangle +
  \sum_{r\neq s} V_{ir}V_{jr}V_{ks}V_{ls}\langle y_r^2 y_s^2\rangle + \nonumber \\
  & \displaystyle{\sum_{r\neq s}} & V_{ir}V_{js}V_{kr}V_{ls}\langle y_r^2 y_s^2\rangle +
  \sum_{r\neq s} V_{ir}V_{js}V_{ks}V_{lr}\langle y_r^2 y_s^2\rangle.
  \label{fp_sum}
\end{eqnarray}
The insertion of $y_r^4$ under the Gaussian integral gives an additional factor of
$3/[4(s-K\lambda_r)^2]=3/[16K^2(z-\lambda_r/2)^2]$, whereas $y_r^2 y_s^2$ gives
$1/[16K^2(z-\lambda_r/2)(z-\lambda_s/2)]$, so that the diagonal terms left out in
Eq.\ (\ref{fp_sum}) are reinserted:
\begin{eqnarray}
  C_{ijkl} = & \sum_{r,s} &
  (V_{ir}V_{jr}V_{ks}V_{ls}+V_{ir}V_{js}V_{kr}V_{ls}\nonumber \\
  & & +V_{ir}V_{js}V_{ks}V_{lr})\;\langle y_r^2 y_s^2\rangle
  \label{spher_fourer}
\end{eqnarray}
Now performing the $z$-integration of Eq.\ (\ref{spher_part}) in the saddle point
limit $N\rightarrow\infty$ is equivalent to just inserting the saddle point value
$z=z_s$ for the factors given above, whenever a normal saddle point exists. As Berlin
and Kac have shown, this is the case for all finite temperatures in one and two
dimensions and for $T\ge T_c$ in three dimensions; in the low-temperature phase, the
saddle point ``sticks'' to its value for $T=T_c$. Then, the four-point function
simply factorizes, so that, comparing Eq.\ (\ref{spher_fourer}) to the expression
Eq.\ (\ref{spher_two}) for the two-point function it is clear that:
\begin{equation}
  C_{ijkl}=C_{ij}C_{kl}+C_{ik}C_{jl}+C_{il}C_{jk},
  \label{wick}
\end{equation}
and, finally, considering the connected energy-energy correlation function, one has:
\begin{eqnarray}
  & C_{i\,i+1\;j\,j+1}-C_{i\,i+1}^2 & = C_{i\,j}C_{i+1\,j+1}+C_{i\,j+1}C_{i+1\,j} \nonumber \\
  & & \longrightarrow 2C_{i\,j}^2,\; |j-i|\rightarrow\infty,
\end{eqnarray}
so that the energy-energy correlation function is trivially related to the spin-spin
correlation function. Note that Eq.\ (\ref{wick}) would follow from Wicks's Lemma for
the Gaussian model. This especially confirms the factor-two relation
$x_\epsilon/x_\sigma=2$ between the corresponding scaling dimensions derived by
Henkel using transfer matrices \cite{henkel:88a}. The factorization property can also
be seen in the grand-canonical formulation of the spherical model, the ``mean''
spherical model \cite{lewis:52a}, where the hard constraint Eq.\ (\ref{constraint})
is being replaced by its thermodynamical average, so that one can leave out the
problematic $z$-integration above. There has been some debate over the coincidence of
the thermodynamic limit of the two models, which is now believed to be settled
\cite{yan:65a}.

\section{Energy-energy correlation function in the limit of infinite spin dimensionality}

The treatment of the partition function of the O($n$) model in the
$n\rightarrow\infty$ limit is quite analogous to that of the spherical model, cp.\ 
\cite{stanley:68a}. For the comparison of the $n\rightarrow\infty$ limit with the
spherical model the constraint $\bm{\sigma}_i\cdot\bm{\sigma}_i=1$ of Eq.\ 
(\ref{Hamilton}) has to be replaced by $\bm{\sigma}_i\cdot\bm{\sigma}_i=n$. We
write the partition function of the model as:
\begin{eqnarray}
   & \displaystyle{Z_N^{(n)}(K)={A_N^{(n)}}^{-1}\int\cdots\int{\rm d}\sigma_1^{(1)}\cdots{\rm
     d}\sigma_N^{(n)}\prod_j\delta(n-\bm{\sigma}_j^2)} & \nonumber \\
   & \displaystyle{\times \exp[K\sum_{\langle ij\rangle}\sum_\nu\sigma_i^{(\nu)}\sigma_j^{(\nu)}]}, &
\label{on_part}
\end{eqnarray}
where $A_N^{(n)}$ ensures the correct normalization. Rewriting the
$\delta$-constraints to the Fourier representation, one now has to introduce $N$
variables $\left\{t_i\right\}$, arriving at:
\begin{eqnarray}
  & \displaystyle{Z_N^{(n)}(K)={A_N^{(n)}}^{-1}{\left(\frac{K}{2\pi i}\right)}^N\int_{-\infty}^{+\infty}
    \cdots\int_{-\infty}^{+\infty}{\rm d}\sigma_1^{(1)}\cdots{\rm d}\sigma_N^{(n)}} &
  \nonumber \\
  & \displaystyle{\times\int_{-i\infty}^{+i\infty}\cdots\int_{-i\infty}^{+i\infty}
    {\rm d}t_1\cdots{\rm d}t_N\,\exp(Kn\sum_i t_i)} & \nonumber \\
  & \displaystyle{\times \prod_{\nu=1}^n\exp(-K\sum_i t_i{\sigma_i^{(\nu)}}^2+
    K\sum_{\langle ij\rangle}\sigma_i^{(\nu)}\sigma_j^{(\nu)})}. & 
\end{eqnarray}
Interchanging the order of integrations one is again left with integrals of Gaussian
type that are easily solved transforming the spin variables orthogonally according to
$\sigma_i^{(\nu)}=\sum_j V_{ij} y_j^{(\nu)}$. Note that the transformation is
symmetric in the component index $\nu$ of the spins. The calculation is given in more
detail for the case of a one-dimensional chain below. Here, we again consider the
relation between two-point and four-point correlation functions. We take the
two-point function to be
\begin{equation}
  C_{ij}\equiv \frac{1}{n}\langle\bm{\sigma}_i\cdot\bm{\sigma}_j\rangle=
  \langle\sigma_i^{(\nu)}\sigma_j^{(\nu)}\rangle,
  \label{two_unbroken}
\end{equation}
where the last equation for any $\nu=1,\ldots,n$ follows from the O($n$) symmetry of
the model in the unbroken, high-temperature phase.  Using the same arguments of
Gaussian integration as for the case of the spherical model, the four-point function:
\begin{equation}
  C_{ijkl}\equiv \frac{1}{n^2}\langle(\bm{\sigma}_i\cdot\bm{\sigma}_j)
  (\bm{\sigma}_k\cdot\bm{\sigma}_l)\rangle,
\end{equation}
again decomposes in terms of the diagonal variables $y_i^{(\nu)}$ as:
\begin{eqnarray}
  C_{ijkl} & = & \displaystyle{\frac{1}{n^2}\sum_{r,t,\mu,\nu}V_{ri}V_{rj}V_{tk}V_{tl}\,
    \langle{y_r^{(\mu)}}^2{y_t^{(\nu)}}^2\rangle} \nonumber \\
  & & \displaystyle{+\frac{1}{n^2}\sum_{r,s,\mu}V_{ri}V_{sj}V_{rk}V_{sl}\,
    \langle{y_r^{(\mu)}}^2{y_s^{(\mu)}}^2\rangle} \nonumber \\
  & & \displaystyle{+\frac{1}{n^2}\sum_{r,s,\mu}V_{ri}V_{sj}V_{sk}V_{rl}\,
    \langle{y_r^{(\mu)}}^2{y_s^{(\mu)}}^2\rangle}.
\end{eqnarray}
In the saddle point limit, which now corresponds to $n\rightarrow\infty$, this
expression factorizes in terms of two-point functions as:
\begin{equation}
  C_{ijkl}=C_{ij}C_{kl}+\frac{1}{n}C_{ik}C_{jl}+\frac{1}{n}C_{il}C_{jk},
  \label{four_unbroken}
\end{equation}
so that the ``mixed'' terms are suppressed with $1/n$. This asymmetry results from
the preset pairing of the spin component indices $\mu$ and $\nu$ in the four-point
function. As a consequence, the connected part of the energy-energy correlation
function:
\begin{equation}
  C_{i\,i+1\;j\,j+1}-C_{i\,i+1}^2 = \frac{1}{n}C_{i\,j}C_{i+1\,j+1}+
  \frac{1}{n}C_{i\,j+1}C_{j\,i+1},
\end{equation}
vanishes in the first-order saddle-point approximation. Thus, any non-vanishing
contributions that are to be expected from our numerical results, have to come from
sub-leading terms in the steepest-descent expansion. The correspondence of the
$n\rightarrow\infty$ limit to the spherical model seems only to hold to leading order
of the saddle-point approximation.

In the broken, low-temperature phase Eq.\ (\ref{two_unbroken}) has to be replaced by
\begin{equation}
  C_{ij}=\frac{1}{n}\langle\bm{\sigma}_i\cdot\bm{\sigma}_j\rangle\le
   \max_\nu\,\langle\sigma_i^{(\nu)}\sigma_j^{(\nu)}\rangle\equiv C_{ij}^{\rm max},
\end{equation}
so that the factorization property of the four-point function Eq.\
(\ref{four_unbroken}) becomes
\begin{equation}
   C_{ijkl}\le C_{ij}C_{kl}+\frac{1}{n}C_{ik}^{\rm max}C_{jl}^{\rm
   max}+\frac{1}{n}C_{il}^{\rm max}C_{jk}^{\rm max},
\end{equation}
and again the connected part of the energy-energy correlation function is $O(1/n)$,
vanishing in the first-order saddle-point limit.

For the case of an one-dimensional lattice the first-order saddle-point approximation
is exact as can be checked by explicit calculation. Consider an open chain of O($n$)
spins\footnote{Considering a {\em closed\/} chain is technically much more intricate,
  cp.\ \cite{thompson:domb}, but, of course, gives the same results in the
  thermodynamic limit $N\rightarrow\infty$.}. The partition function is given by the
general expression Eq.\ (\ref{on_part}) with the nearest-neighbor sum $\sum_{\langle
  ij\rangle}\bm{\sigma}_i\cdot\bm{\sigma}_j$ replaced by the one-dimensional
expression $\sum_{i}\bm{\sigma}_i\cdot\bm{\sigma}_{i+1}$. Following Stanley
\cite{stanley:69a}, we factor out the integration over the last spin $\bm{\sigma}_N$,
which has the form:
\begin{eqnarray}
  & \displaystyle{{\cal Z}^{(n)}(K)=\frac{K}{2\pi i}\int\cdots\int{\rm d}\sigma^{(1)}\cdots{\rm
      d}\sigma^{(n)}
  \int_{-i\infty}^{+i\infty}{\rm d}u} & \nonumber \\
  & \times\displaystyle{\exp[uK(n-\sum_\nu{\sigma^{(\nu)}}^2)]\,\exp[K\sum_\nu c_\nu\sigma^{(\nu)}]}, &
\end{eqnarray}
where $c_\nu\equiv\sigma_{N-1}^{(\nu)}$. Inserting the unity factor
$\exp[K\alpha_0(n-\sum_\nu{\sigma^{(\nu)}}^2)]$ and choosing $\alpha_0$ sufficiently
large to exclude the singularities, one has:
\begin{eqnarray}
  \displaystyle{{\cal Z}^{(n)}(K)} & = & \frac{K}{2\pi
      i}\int_{\alpha_0-i\infty}^{\alpha_0+i\infty}{\rm d}v\,
    e^{vKn}\prod_\nu\int{\rm d}\sigma^{(\nu)}
  \nonumber \\
  & &  \displaystyle{\times \exp[-K(v{\sigma^{(\nu)}}^2-c_\nu\sigma^{(\nu)})]},
\end{eqnarray}
where $v\equiv u+\alpha_0$. Square completion and a change of variables $w=2v$ gives:
\begin{eqnarray}
  \displaystyle{{\cal Z}^{(n)}(K)} & = & {\left(\frac{2\pi}{K}\right)}^{n/2}\frac{K}{4\pi
      i}\int_{2\alpha_0-i\infty}^{2\alpha_0+i\infty}
    {\rm d}w \nonumber \\ 
  & & \times\displaystyle{\exp[\frac{1}{2}nK(w+1/w)]w^{-n/2}} \nonumber \\
   & = & \frac{1}{2}K(2\pi/K)^{n/2}I_{n/2-1}(nK),
\label{stan_gauss}
\end{eqnarray}
which is an integral representation of the modified Bessel function of the first
kind. Thus, the spin integrations can be done successively, the full partition
function being given by:
\begin{equation}
  Z_N^{(n)}(K)=[(nK/2)^{1-n/2}\Gamma(n/2)I_{n/2-1}(nK)]^{N-1},
\end{equation}
where the $\Gamma$ function enters through the normalization factor
${A_N^{(n)}}^{-1}$ and the last integration, which corresponds to ${\cal
  Z}^{(n)}(0)$.  Considering the two-point function, an additional factor
$\bm{\sigma}_i\cdot\bm{\sigma}_j$, $i<j$, is inserted in the integrand of Eq.
(\ref{on_part}).  Again starting the integration with the last spin $\bm{\sigma}_N$,
the first $N-j$ integrations are unaltered. The integration over $\bm{\sigma}_j$
gives additional factors of $c_\nu/2v$ from the Gaussian integration Eq.\ 
(\ref{stan_gauss}), where now $c_\nu\equiv\sigma_{j-1}^{(\nu)}$, so that one is left
with
\begin{equation}
    \tilde{{\cal Z}}^{(n)}(K) = \frac{1}{2}K{\left(\frac{2\pi}{K}\right)}^{n/2}
    I_{n/2}(nK)\sum_\nu\sigma^{(\nu)}_i c_\nu, 
\end{equation}
and the form of the integrand for the next integrations is unchanged. The integration
over $\bm{\sigma}_i$ adds a factor of $n$ since $c_\nu$ above becomes
$\sigma^{(\nu)}_i$ and $\sum_\nu\sigma_i^{(\nu)}\cdot\sigma_i^{(\nu)}=n$, followed by
another $i-1$ integrations of the partition-function type. With $u\equiv
u(nK)=I_{n/2}(nK)$ and $v\equiv v(nK)=I_{n/2-1}(nK)$ one arrives at:
\begin{equation}
  \frac{1}{n}\langle\bm{\sigma}_i\cdot\bm{\sigma}_j\rangle =
  \frac{v^{N-1+i-j}u^{j-i}}{v^{N-1}}=(u/v)^{j-i}.
\end{equation}
From this it is straightforward to derive the form of the four-point
function by analogy:
\begin{eqnarray}
  \displaystyle{\frac{1}{n^2}\langle(\bm{\sigma}_i\cdot\bm{\sigma}_{j})
    (\bm{\sigma}_{k}\cdot\bm{\sigma}_{l})\rangle} & = &
  v^{N-l}u^{l-k}v^{k-j}u^{j-i}v^{i-1}v^{1-N} \nonumber \\ & = & (u/v)^{(l-k)+(j-i)},
\end{eqnarray}
where $i<j<k<l$ is understood. For the special case of energy-energy correlations one
has:
\begin{equation}
  \frac{1}{n^2}\langle(\bm{\sigma}_i\cdot\bm{\sigma}_{i+1})
  (\bm{\sigma}_{j}\cdot\bm{\sigma}_{j+1})\rangle = (u/v)^2,
\end{equation}
which does not depend on the distance $|j-i|$. Hence the connected energy-energy
correlation function vanishes exactly even for finite $n$ in one dimension.  The
$n\rightarrow\infty$ limit of this expression is given by:
\begin{equation}
  \frac{1}{n^2}\langle(\bm{\sigma}_i\cdot\bm{\sigma}_{i+1})
  (\bm{\sigma}_{j}\cdot\bm{\sigma}_{j+1})\rangle =
  \frac{4K^2}{[1+\sqrt{1+(2K)^2}]^2}.
\end{equation}

%\bibliography{general}

\end{document}